\documentclass[twocolumn,preprintnumbers,amsmath,amssymb]{revtex4}

	\usepackage{amsmath}
	\usepackage{makeidx}
	\usepackage{amsfonts}
	\usepackage[ansinew]{inputenc}
	\usepackage[usenames,dvipsnames]{pstricks}
	\usepackage{subfigure}
	\usepackage{epsfig}
	\usepackage{graphicx}
	\usepackage{dcolumn}
	\usepackage{bm}
	\usepackage[colorlinks,hyperindex]{hyperref}
	\usepackage{epstopdf}	
	\usepackage{float}
	\usepackage{perpage}
	\usepackage{lineno}
	\usepackage{booktabs}

	\MakeSorted{figure}
	\MakeSorted{table}
	\hypersetup
	{
		colorlinks,%
		citecolor=black,%
		linkcolor=black,%
		urlcolor=black,%
	}

	\newcommand{\um}[1]{\mathrm{#1}}

\makeindex

\begin{document}

\title{Impulsive stimulated scatttering signal in supercooled liquids with Debye or Havriliak-Negami relaxation of the specific heat capacity and thermal expansion coefficient}

\author{Marco Gandolfi}
\email{marco.gandolfi@ino.cnr.it}
\affiliation{CNR-INO (National Institute of Optics), Via Branze 45, Brescia, Italy}
\affiliation{Department of Information Engineering, University of Brescia, Via Branze 38, Brescia, Italy}
\affiliation{Laboratory of Soft Matter and Biophysics, Department of Physics and Astronomy, KU Leuven, Celestijnenlaan 200D, B-3001 Leuven, Belgium}
\affiliation{Dipartimento di Matematica e Fisica, Universit\`{a} Cattolica del Sacro Cuore, Via Musei 41, 25121 Brescia, Italy}
\affiliation{Interdisciplinary Laboratories for Advanced Materials Physics (I-LAMP), Via Musei 41, 25121 Brescia, Italy}

\author{Liwang Liu}
\email{liwang.liu@kuleuven.be}
\affiliation{Laboratory of Soft Matter and Biophysics, Department of Physics and Astronomy, KU Leuven, Celestijnenlaan 200D, B-3001 Leuven, Belgium}

\author{Pengfei Zhang}
\affiliation{Laboratory of Soft Matter and Biophysics, Department of Physics and Astronomy, KU Leuven, Celestijnenlaan 200D, B-3001 Leuven, Belgium}

\author{Robbe Salenbien}
\affiliation{VITO, Boeretang 200, Mol 2400, Belgium}
\affiliation{EnergyVille, Energyville I, Thor Park 3800
Genk 3600, Belgium}

\author {Francesco Banfi}
\affiliation{FemtoNanoOptics group, Universit\'{e} de Lyon, CNRS, Universit\'{e} Claude Bernard Lyon 1, Institut Lumi\`{e}re Mati\`{e}re, F-69622 Villeurbanne, France}
\affiliation{Interdisciplinary Laboratories for Advanced Materials Physics (I-LAMP), Via Musei 41, 25121 Brescia, Italy}

\author{Christ Glorieux}
\affiliation{Laboratory of Soft Matter and Biophysics, Department of Physics and Astronomy, KU Leuven, Celestijnenlaan 200D, B-3001 Leuven, Belgium }

\begin{abstract}
A generalized physical model is introduced to describe the impulsive stimulated scattering (ISS) response of relaxing systems to photothermal excitation in a periodical grating geometry. The proposed approach starts from Debye and Havriliak-Negami expressions for both the frequency-dependent heat capacity, $C(\omega)$, and thermal expansion coefficient, $\gamma(\omega)$. Simulations are carried out on glycerol to test and compare the developed models with the  existing semi-empirical model \cite{yang1995impulsive1}. Debye behavior of the specific heat capacity is shown to be compatible with a two-temperature scenario, in which, in addition to the classical, experimentally observable temperature that characterizes the distribution of the system over vibrational energy states, a second temperature characterizes the state of the configurational energy landscape. The models  here developed have been applied for the interpretation of the experimental ISS signals of supercooled glycerol, illustrating simultaneous and separate assessment of $C(\omega)$ and $\gamma(\omega)$ up to sub-100 MHz from thermoelastic transients.

\end{abstract}

\maketitle

\section{Introduction\label{Introduction}}
The intriguing behavior of glass-forming liquids is attracting continued interest from many researchers \cite{bapst2020unveiling,jensen2018slow,hecksher2017toward,klieber2013mechanical,blazhnov2004temperature,niss2018perspective,klieber2015nonlinear,gundermann2011predicting}.
By virtue of its ability to simultaneously probe multiple relaxation processes (thermal expansion, acoustic and even orientational response \cite{glorieux2002thermal,silence1992structural}), the use of impulsive stimulated scattering (ISS) in a periodical grating geometry has been successful in obtaining new insights from the thermoelastic response (measured via the accompaning coherent diffraction of a probe laser beam) to impulsive photothermal excitation of different glassformers \cite{yang1995t,yang1995impulsive2,paolucci2000impulsive,halalay1992liquid,halalay1992time,silence1992structural,silence1990impulsive}.   
Standard thermo-mechanical modelling, based on the assumption of frequency independent or non-relaxing heat capacity and thermal expansion coefficient, have been shown not to be adequate to characterize the dynamics triggered in an ISS experiment, especially for viscous systems.
Along with the first experimental ISS results, a semi-empirical model (SEM) \cite{yang1995impulsive1}, relying on a stretched-exponential function to describe the nontrivial initial thermal expansion rise of the ISS signal, has proved to be effective to describe the ISS response of glycerol, salol, and oil DC705 \cite{yang1995t,yang1995impulsive1,yang1995impulsive2,paolucci2000impulsive}.\\
Inspired by successful descriptions in literature of experimental results for the temperature response to heating \cite{birge1985specific,bentefour2003broadband,bentefour2004thermal,niss2012dynamic} by invoking a frequency dependent heat capacity $C$, and indications for a frequency dependent thermal expansion coefficient $\gamma$ \cite{blazhnov2004temperature}, here, we derive analytically a generalized ISS model to take into account the relaxation of $C$ and $\gamma$, which are not implicitly considered in the SEM.
We start from frequency domain versions of the thermal diffusion equation and the thermoelastic equation and we impose a frequency dependent heat capacity and thermal expansion coefficient according to Debye and Havriliak-Negami (HN) relaxation models.
We then investigate to what extent the introduced physical model is consistent with the empirical model by conducting a case study on ISS results of glycerol reported in Refs. \cite{paolucci2000impulsive,Liu2021}.
We also present an interpretation of the Debye assumption for the frequency dependent heat capacity and thermal expansion coefficient in the framework of a two-temperature model (TTM). Furthermore, a set of experimentally recorded ISS signals of a supercooled glycerol is analysed with the developed models to extract $C(\omega)$ and $\gamma(\omega)$ up to sub-100 MHz. This largely extends the upper limit of the previously accessible bandwidth, 100 kHz \cite{bentefour2003broadband} and 1 Hz \cite{niss2012dynamic} respectively for the spectroscopy of \textit{C} and $\gamma$, enabling the comparison of fragility by thermal, mechanical, and dielectric susceptibilities in a broader frequency/temperature range. \\
The manuscript is organized as follows:
In Section \ref{Sec:Temperature} we present analytical expressions for the temperature response to impulse photothermal excitation in a grating geometry in two scenarios: frequency independent and frequency dependent (according to Debye and HN functions) heat capacity.
In Section \ref{Sec:displacement} a continuum mechanics thermoelastic model is used to calculate the response of the material strain to photothermal excitation, by considering the temperature change as source term in the equation of motion, in which Debye and HN relaxation behavior of thermal expansion is coupled.
A comparison between results obtained by the newly proposed approach and simulations by the empirical model for literature values on glycerol \cite{paolucci2000impulsive} is presented in Section \ref{Sec:comparison_with_liter}. In Section \ref{Sec:two_temp_model} the compatibility between Debye frequency dependence of the heat capacity and of thermal expansion coefficient and the TTM is verified. Finally, in Section \ref{experiment}, we apply the developed models to the concrete case of the experimental ISS signals recorded on a supercooled glycerol.\\
The present work accompanies the results presented in Ref. \cite{Liu2021}, in which the thermal relaxation dynamics of glycerol is investigated by a combination of ISS and thermal lens spectroscopy \cite{ThermalLens}.\\

\section{Temperature response to impulsive photothermal excitation in a periodical grating geometry}
\label{Sec:Temperature}

\subsection{Scenario with frequency-independent heat capacity \label{Subsec:Frequency_independent_heat_capacity}}
In this section we calculate the temperature evolution of a system that is subject to impulsive photothermal excitation generating a transient thermal grating (TTG). For now, we assume that the heat capacity is frequency independent.\\
The starting point is the thermal diffusion equation in the temperature $T$ for a 1D infinite geometry \cite{gandolfi2019accessing}:
\begin{equation}
\frac{\partial^2 T}{\partial x^2 }-\frac{\rho C}{\kappa_T} \frac{\partial T}{\partial t}=-\frac{Q(x,t)}{\kappa_T},
\label{diffusion_equation2}
\end{equation}
where $\rho$  (kg m$^{-3}$), $\kappa_T$ (W m$^{-1}$ K$^{-1}$) and $C$  (J kg$^{-1}$ K$^{-1}$)) are the mass density, the thermal conductivity  and the frequency-independent heat capacity per unit mass, while $Q(x,t)$ (W m$^{-3}$) is the heat source. In ISS experiments, the heat input is impulsive in time and periodical in space: 
\begin{equation}
Q(x,t)=Q_0\cos(qx)\delta(t),
\label{Q_x_t}
\end{equation}
where $Q_0$ (J m$^{-3}$) is the supplied heat density and $q$ (m$^{-1}$) is the wavenumber, defined as $2\pi$ times the inverse of spatial period of periodical light intensity pattern. \\
Prior to excitation, the system is at equilibrium at constant temperature $T_0$.\\
After taking a Fourier transform, the following frequency domain expression is obtained:
\begin{equation}
\frac{\partial^2 \tilde{T}}{\partial x^2 }-i\omega\frac{\rho C}{\kappa_T} \tilde{T}=-\frac{1}{\kappa_T}\tilde{Q}(x,\omega),
\label{diffusion_equation_in_freq_domain_AA}
\end{equation}
and the solution for temperature field reads:
\begin{equation}
\tilde{T}(x,\omega)=T_0\delta(\omega)+\frac{Q_0}{2\pi i \rho C( \omega-i\alpha q^2)}\cos(qx),
\label{temperature_in_freq}
\end{equation}
where $\alpha=\kappa_T/(\rho C)$ (m$^2$ s) is the thermal diffusivity.
By taking an inverse Fourier transform, the following expression is obtained for the temperature evolution:
\begin{equation}
T(x,t)=T_0+\frac{Q_0}{\rho C}\cos(qx)\exp\left(-\alpha q^2 t\right)\theta(t),
\label{temperature_in_time}
\end{equation}
where $\theta(t)$ is the Heaviside step function.\\

\subsection{Scenario with frequency dependent heat capacity}
\label{Subse:Frequ-dependent heat capacity}
\subsubsection{Debye model \label{Subsum:Debye}}
The following Debye expression for the frequency dependent heat capacity  \cite{fivez2011dynamics}:
\begin{equation}
C(\omega)=C_\infty+\frac{\Delta C}{1+i\omega \tau_C}=C_\infty+\frac{\Delta C}{1+i\frac{\omega}{\omega_C}},
\label{C_omega}
\end{equation}
with $C_\infty$ the part of the heat capacity related to the higher frequency response, or, in time domain, the instantaneous response of the temperature to impulsive heating. $\Delta C$ is the additional part of the heat capacity that determines the reduction of the temperature response at low frequencies (lower than the relaxation frequency  $\omega_C=\tau_C^{-1}$), or, in time domain, at times longer than the relaxation time $\tau_C$.

Upon substitution of the expression for $C(\omega)$ into Eq. \ref{diffusion_equation_in_freq_domain_AA} we get the following differential equation:
\begin{equation}
\frac{\partial^2 \tilde{T}}{\partial x^2 }-i\omega\frac{\rho}{\kappa_T} \left(C_\infty+\frac{\Delta C}{1+i{\frac{\omega}{\omega_C} }}\right)\tilde{T}=-\frac{1}{\kappa_T}\tilde{Q}(x,\omega).
\label{diffusion_equation_in_freq_C_omega}
\end{equation}
Inserting the expression for the heat source $\tilde{Q}(x,\omega)$ (obtained transforming in frequency domain Eq. \ref{Q_x_t}), we obtain the following solution:
\begin{equation}
\tilde{T}(x,\omega)=T_0\delta(\omega)-\frac{iQ_0\left(\omega-i\omega_C\right)}{2\pi \rho C_\infty\left(\omega-\omega_1\right)\left(\omega-\omega_2\right)}\cos(qx),
\label{temperature_in_freq_C_omega}
\end{equation}
with $\alpha_\infty=\kappa_T/(\rho C_\infty)$ the high frequency limit of the thermal diffusivity and the frequences
\begin{widetext}
\begin{eqnarray}
\omega_1=\frac{i}{2}\left\{\left[\alpha_\infty q^2+\omega_C\left(1+\frac{\Delta C}{C_\infty}\right)\right]-\sqrt{\left[\alpha_\infty q^2+\omega_C\left(1+\frac{\Delta C}{C_\infty}\right)\right]^2-4\alpha_\infty q^2\omega_C}\right\}, \label{omega_1} \label{omega1_def}\\
\nonumber\\
\omega_2=\frac{i}{2}\left\{\left[\alpha_\infty q^2+\omega_C\left(1+\frac{\Delta C}{C_\infty}\right)\right]+\sqrt{\left[\alpha_\infty q^2+\omega_C\left(1+\frac{\Delta C}{C_\infty}\right)\right]^2-4\alpha_\infty q^2\omega_C}\right\}, \label{omega2_def}\\
\nonumber
\end{eqnarray}
The expression for the temperature in time domain can be obtained by applying the inverse Fourier transform to Eq. \ref{temperature_in_freq_C_omega} (see Appendix \ref{appen:res_th} for more details).
Thus, we obtain:
$$T(x,t)=T_0+\frac{Q_0}{\rho C_\infty}\cos(qx)\times\left[\frac{\left(\omega_1-i\omega_C\right)}{\left(\omega_1-\omega_2\right)}\exp\left(i\omega_1 t\right)+\frac{\left(\omega_2-i\omega_C\right)}{\left(\omega_2-\omega_1\right)}\exp\left(i\omega_2 t\right)\right]\theta(t)=$$

$$=T_0+\frac{Q_0}{\rho C_\infty}\cos(qx)\exp\left\{-\frac{t}{2}\left[\alpha_\infty q^2+\omega_C\left(1+\frac{\Delta C}{C_\infty}\right)\right]\right\}\times$$

$$\times\left\{\cosh\left({t\over2}\sqrt{\left[\alpha_\infty q^2+\omega_C\left(1+\frac{\Delta C}{C_\infty}\right)\right]^2-4\alpha_\infty q^2\omega_C}\right)\right.+$$

\begin{equation}
-\frac{\left[\alpha_\infty q^2+\omega_C\left(\frac{\Delta C}{C_\infty}-1\right)\right]}{\sqrt{\left[\alpha_\infty q^2+\omega_C\left(1+\frac{\Delta C}{C_\infty}\right)\right]^2-4\alpha_\infty q^2\omega_C}}\left.\sinh\left({t\over2}\sqrt{\left[\alpha_\infty q^2+\omega_C\left(1+\frac{\Delta C}{C_\infty}\right)\right]^2-4\alpha_\infty q^2\omega_C}\right)\right\}\theta(t).
\label{T_t_expression_final}
\end{equation}

\subsubsection{Havriliak-Negami model \label{Subsum:Havriliak}}
Simple Debye relaxation behavior has turned out not to be a fully adequate description for the dynamic behavior of many glass-forming materials.
By virtue of two additional model parameters $a_C$ and $b_C$, the generalized \textit{Havriliak-Negami} (HN) model \cite{havriliak1966complex}:
\begin{equation}
C(\omega)=C_\infty+\frac{\Delta C}{\left[1+\left(i\omega \tau_C\right)^{a_C}\right]^{b_C}}=C_\infty+\frac{\Delta C}{\left[1+\left(i\frac{\omega}{\omega_C}\right)^{a_C}\right]^{b_C}},
\label{C_omega_HN}
\end{equation}
 has been shown to be more effective (note that the HN model tends to the Debye model when $a_C=b_C=1$).\\

In this case, the temperature response in the frequency domain is given by:

\begin{equation}
\tilde{T}(x,\omega)=T_0\delta(\omega)-\frac{iQ_0\left[\omega_C^{a_C}+(i\omega)^{a_C}\right]^{b_C}\cos(qx)}{2\pi \rho C_\infty\left\{\omega\left[\omega_C^{a_C}+(i\omega)^{a_C}\right]^{b_C}+\frac{\Delta C}{C_\infty}\omega\omega_C^{a_Cb_C}-i\alpha_\infty q^2\left[\omega_C^{a_C}+(i\omega)^{a_C}\right]^{b_C}\right\}}.
\label{temperature_in_freq_C_omega_HN}
\end{equation}
As the exponents $a_C$ and $b_C$ are typically non-integer, analytical derivation of the inverse Fourier transform of the latter expression is cumbersome. Therefore, in this work, this inverse Fourier was performed numerically.\clearpage
\end{widetext}

\section{ISS Signal}
\label{Sec:displacement}

\subsection{Constitutive equation}
\label{SubSec:Constitutive_Eq}
Impulsive stimulated scattering occurs due to coherent diffraction of a probe beam that trespasses a medium of interest in which, via its relation with the refractive index, a spatially periodic strain pattern is present. The ISS signal is therefore proportional to the magnitude of the strain grating in the medium.
In this subsection we derive, for different scenarios for the relaxation behavior an expression for the displacement and strain.
We assume that the viscoelastic behavior of the material can be described by the Kelvin-Voigt model, corresponding with a lumped model containing a spring and a dashpot in parallel (as described on page 87 of Ref. \cite{auld1973acoustic}).\\
Under this assumption, the constitutive equations are:
\begin{equation}
\left\{\begin{array}{l}
\displaystyle{\rho \frac{\partial^2 \mathbf{u}}{\partial t^2}=\nabla\cdot \sigma}\\
\\
\displaystyle{\sigma=\mathbf{C}\varepsilon+\eta\frac{\partial \varepsilon}{\partial t},}
\end{array}\right.
\label{constitutive_relations}
\end{equation}
where $\mathbf{u}$ (m) is the displacement, $\sigma$ (Pa) is the stress, $\mathbf{C}$ (Pa) is the stiffness matrix and $\eta$ (Pa s) is the viscosity tensor. The strain $\varepsilon$ can be written as:
\begin{equation}
\varepsilon=\nabla_S\mathbf{u}-\gamma_M \Delta T,
\label{definition_of_strain_C}
\end{equation}
where $\nabla_S\mathbf{u}=\frac{\nabla \mathbf{u}+\nabla^T\mathbf{u}}{2}$, $\gamma_M$ [$K^{-1}$] is the matrix of linear expansion and $\Delta T$ is temperature variation that drives the mechanics \cite{gandolfi2020optical}. This approach is in agreement with Green-Lindsay theory for thermoviscoelastic media \cite{mukhopadhyay1999relaxation,othman2012fundamental}.\\
We write the viscoelastic tensor as $\eta=\tau_\eta \mathbf{C}$
\footnote{As indicated on page 88 of Ref. \cite{auld1973acoustic}, the viscosity tensor has same form of the stiffness matrix.
Hence, we can write viscosity tensor as $\eta=\tau_\eta C$, where $C$ is the stiffness matrix \cite{mukhopadhyay1999relaxation,othman2012fundamental}.}, where $\tau_\eta$ representes the damping time and we assume that the medium is homogeneous and isotropic. Upon these assumptions, the equation ruling the displacement reads:
\vspace{3cm}
\begin{widetext}
\begin{equation}
\frac{\partial^2 u_x}{\partial t^2}=
c_L^2\left(1+\tau_\eta\frac{\partial}{\partial t}\right)\frac{\partial^2 u_x}{\partial x^2}-(3c_L^2-4c_T^2)\gamma\left(1+\tau_\eta\frac{\partial}{\partial t}\right)\frac{\partial T}{\partial x},
\label{Expansion_isotropic}
\end{equation}
where $c_L=\sqrt{(\lambda+2\mu)/\rho}$ and $c_T=\sqrt{\mu/\rho}$ are the longitudinal and transverse velocities (m/s), with $\lambda$ (Pa) and $\mu$ (Pa) the two Lam\'{e} coefficients. $\gamma$ is the linear expansion coefficient.
\end{widetext}

\subsection{ISS response in case of frequency independent heat capacity and thermal expansion}
\label{SubSec:Freq_ind_u}
In this subsection we derive, for a non-relaxing medium, a general expression for the displacement occurring when the system is excited by a transient optical grating. \\
To do this, we apply the temporal Fourier transform to Eq. \ref{Expansion_isotropic} to get:
$$-\omega^2 \tilde{u}_x=
c_L^2(1+i\omega\tau_\eta)\frac{\partial^2 \tilde{u}_x}{\partial x^2}+$$
\begin{equation}
-\left(3c_L^2-4c_T^2\right)\left(1+i\omega \tau_\eta\right)\gamma\frac{\partial \tilde{T}}{ \partial x}.
\label{Expansion_isotropic_FT}
\end{equation}
By defining:
\begin{equation}
c^2(\omega)=c_L^2(1+i\omega\tau_\eta)
\label{def_c}
\end{equation}
and
\begin{equation}
\xi(\omega)=3-4\frac{c_T^2}{ c_L^2},
\label{def_xi}
\end{equation}
we can write Eq. \ref{Expansion_isotropic_FT} in the more compact form:
\begin{equation}
\frac{\partial^2 \tilde{u}_x}{\partial x^2}+\frac{\omega^2}{c^2(\omega)} \tilde{u}_x=\xi(\omega)\gamma\frac{\partial \tilde{T}}{ \partial x}.
\label{Expansion_isotropic_FT_compact}
\end{equation}
In order calculate the displacement occurring due to the TTG excitation, we use the solution for the temperature in frequency domain, as derived in Section \ref{temperature_in_freq}:

\begin{equation}
\frac{\partial^2 \tilde{u}_x}{\partial x^2}+\frac{\omega^2}{c^2(\omega)} \tilde{u}_x=Z(\omega) \sin(qx),
\label{Expansion_isotropic_FT_compact2}
\end{equation}
where
\begin{equation}
Z(\omega)=-\frac{qQ_0\xi(\omega)\gamma}{2\pi i \rho C( \omega-i\alpha q^2)}.
\label{def_Z}
\end{equation}
 The general solution of Eq. \ref{Expansion_isotropic_FT_compact2} is $\tilde{u}(x,\omega)=z(x,\omega)+z_p(x,\omega)$, where $z_p(x,\omega)$ is a particular solution of Eq. \ref{Expansion_isotropic_FT_compact2}, while $z(x,\omega)$ is the solution of the associated homogeneous differential equation.
It can be shown that 
\begin{equation}
z_p(x,\omega)=\frac{Z(\omega)c^2(\omega)}{\omega^2- q^2c^2(\omega)}\sin(qx)
\end{equation}
is a particular solution of Eq. \ref{Expansion_isotropic_FT_compact2}.\\
In order to have the system at rest before the excitation (i.e. $u(x,t)=0$ and $\frac{du}{dt}(x,t)=0$ for negative times) and avoid divergence of the displacement at infinity, we must have $z(x,\omega)=0$ $\forall \omega$.\\
Hence the final solution is:
\begin{equation}
\tilde{u}(x,\omega)=z_p(x,\omega)=\frac{Z(\omega)c^2(\omega)}{\omega^2- q^2c^2(\omega)}\sin(qx)
\end{equation}
Substituting back the expressions for $Z(\omega)$, $\xi(\omega)$ and for $c(\omega)$ in the latter equation we obtain:

\begin{equation}
\tilde{u}(x,\omega)=-\frac{qQ_0\gamma\left(3c_L^2-4c_T^2\right)\left(1+i\omega \tau_\eta\right)}{2\pi i \rho C(\omega-i\alpha q^2)\left(\omega-\omega_3\right)\left(\omega-\omega_4\right)}\sin(qx),
\label{u_omega_substituted2}
\end{equation}
where
\begin{equation}
\omega_3=i\frac{q^2}{2}\left[c_L^2\tau_\eta-\sqrt{c_L^4\tau_\eta^2-\frac{4\rho}{q^2}c_L^2} \right].
\label{def_of_omega3}
\end{equation}
and
\begin{equation}
\omega_4=i\frac{q^2}{2}\left[c_L^2\tau_\eta+\sqrt{c_L^4\tau_\eta^2-\frac{4\rho}{q^2}c_L^2} \right].
\label{def_of_omega4}
\end{equation}

The time domain expression for the displacement can be obtained by applying an inverse Fourier transform to  Eq. \ref{u_omega_substituted2} (see Appendix \ref{appen:res_th}), resulting in:

\begin{widetext}
$$u(x,t)=-\frac{qQ_0\gamma}{\rho C}\sin(qx)\left(3c_L^2-4c_T^2\right)\left\{\frac{1-\alpha q^2 \tau_\eta}{\left(i\alpha q^2-\omega_3\right)\left(i\alpha q^2-\omega_4\right)}\exp\left(-\alpha q^2t\right)\right.+$$
\begin{equation}
\left.+\frac{1+i\omega_3 \tau_\eta}{(\omega_3-i\alpha q^2)\left(\omega_3-\omega_4\right)}\exp\left(i \omega_3 t\right)+\frac{1+i\omega_4 \tau_\eta}{(\omega_4-i\alpha q^2)\left(\omega_4-\omega_3\right)}\exp\left(i \omega_4 t\right)\right\}\theta(t)
\label{u_t_non_dep_on_freq}
\end{equation}

\subsection{ISS response in case of frequency dependent heat capacity and thermal expansion described by the Debye model}
\label{SubSec:Freq_dep_u}
In this section we assume that the heat capacity depends on frequency according to Debye model, in analogy with Subsection \ref{Subsum:Debye}.
Hence, we substitute Eq. \ref{C_omega} into Eq. \ref{u_omega_substituted2} to get:
\begin{equation}
\tilde{u}(x,\omega)=-\frac{qQ_0\gamma\left(3c_L^2-4c_T^2\right)\left(1+i\omega \tau_\eta\right)(\omega-i\omega_C)}{2\pi i \rho C_\infty(\omega-\omega_1)(\omega-\omega_2)\left(\omega-\omega_3\right)\left(\omega-\omega_4\right)}\sin(qx),
\label{u_omega_substituted_C_omega}
\end{equation}
\end{widetext}
where $\omega_1$ and $\omega_2$ are defined in Eq.s \ref{omega1_def} and \ref{omega2_def}.\\
Furthermore, we also assume the linear expansion coefficient to be frequency dependent following the Debye expression:
\begin{equation}
\gamma(\omega)=\gamma_\infty+\frac{\Delta \gamma}{1+i\omega \tau_\gamma}=\gamma_\infty+\frac{\Delta \gamma}{1+i\frac{\omega}{\omega_\gamma}},
\label{gamma_omega}
\end{equation}
where $\gamma_\infty$ and $\Delta \gamma$ represent the instantaneous and additional relaxing contribution of the response of the volume to a temperature change, respectively.
$\omega_\gamma=\tau_\gamma^{-1}$ is the associated relaxation frequency.\\
\\
For the sake of simple analytical treatment, and given that the focus of this work is on the thermal expansion part of the signal and not on the superposed acoustic part, in the following, we neglect the frequency and temperature dependence of the elastic moduli and of the density \cite{jensen2018slow,hecksher2017toward,klieber2013mechanical,blazhnov2004temperature}. \\
With this choice, the equation to be solved reduces to:
\begin{equation}
\frac{\partial^2 \tilde{u}_x}{\partial x^2}+\frac{\omega^2}{c^2(\omega)} \tilde{u}_x=\xi(\omega)\left(\gamma_\infty+\frac{\Delta \gamma}{1+i\frac{\omega}{\omega_\gamma}}\right)\frac{\partial \tilde{T}}{ \partial x}.
\label{Expansion_isotropic_FT_compact_gamma_omega}
\end{equation}

The expression for $\gamma(\omega)$ can be rewritten as:
\begin{equation}
\gamma(\omega)=\gamma_\infty\frac{\omega-i\omega_\gamma\left(1+\frac{\Delta \gamma}{\gamma_\infty}\right)}{\omega-i\omega_\gamma}=\gamma_\infty\frac{\omega-\omega_6}{\omega-\omega_5},
\label{gamma_omega_2}
\end{equation}
where 
\begin{equation}
\omega_5=i\omega_\gamma
\label{omega5_def} 
\end{equation}
and
\begin{equation}
\omega_6=i\omega_\gamma\left(1+\frac{\Delta \gamma}{\gamma_\infty}\right).
\label{omega6_def} 
\end{equation} \vspace{0.5 cm}

Substituting Eq. \ref{gamma_omega_2} into Eq. \ref{u_omega_substituted_C_omega} we get:
\begin{widetext}
\begin{equation}
\tilde{u}_x(x,\omega)=-\frac{qQ_0\gamma_\infty\left(3c_L^2-4c_T^2\right)\left(1+i\omega \tau_\eta\right)(\omega-i\omega_C)(\omega-\omega_6)}{2\pi i \rho C_\infty\prod_{j=1}^{5}(\omega-\omega_j)}\sin(qx).
\label{u_omega_substituted_C_omega2}
\end{equation}
By inverse Fourier inverse transforming the latter expression we obtain the solution for the displacement in time domain:
\begin{equation}
u_x(x,t)=-\left[\frac{qQ_0\gamma_\infty \sin(qx)}{\rho C_\infty}\right]\left(3c_L^2-4c_T^2\right)
\sum_{l=1}^{5}\left\{\left(1+i\omega_l \tau_\eta\right)(\omega_l-i\omega_C)(\omega_l-\omega_6)\left(\prod_{\substack{j=1 \\ j\neq l}}^{5}{1\over\omega_l-\omega_j}\right)\exp(i\omega_l t)\right\}\theta(t).
\label{u_t_substituted_C_omega}
\end{equation}
The ISS signal $U_{ISS}(t)$ is proportional to the amplitude of the strain grating \cite{fivez2011dynamics,yan1987impulsive} $\Delta \rho/\rho$.
Hence, the strain of the 1D displacement pattern equally its spatial derivative, the ISS signal can be derived from Eq. \ref{u_t_substituted_C_omega} as:
\begin{equation}
U_{ISS}(t)\propto \max_{x}\left[\frac{\partial u_{x}(x,t)}{\partial x}\right].
\end{equation}

\subsection{ISS response in case of frequency dependent heat capacity and thermal expansion described by the Havriliak-Negami model}
\label{SubSec:Freq_dep_u_HN}
In the HN scenario for the thermal expansion response,
\begin{equation}
\gamma(\omega)=\gamma_\infty+\frac{\Delta \gamma}{\left[1+\left(i\omega \tau_\gamma\right)^{a_\gamma}\right]^{b_\gamma}}=\gamma_\infty+\frac{\Delta \gamma}{\left[1+\left(i\frac{\omega}{\omega_\gamma}\right)^{a_\gamma}\right]^{b_\gamma}},
\label{gamma_omega_HN}
\end{equation}
with $a_\gamma$ and $b_\gamma$ are additional model parameters.\\
Taking also the heat capacity behavior according to the HN model, as described in Subsection \ref{Subsum:Havriliak} we get:
\begin{equation}
\tilde{u}(x,\omega)=-\frac{qQ_0\gamma_\infty\left(3c_L^2-4c_T^2\right)\left(1+i\omega B_C\tau_\eta\right)\left[\omega_C^{a_C}+(i\omega)^{a_C}\right]^{b_C}\left\{\left[\omega_\gamma^{a_\gamma}+(i\omega)^{a_\gamma}\right]^{b_\gamma}+\frac{\Delta \gamma}{\gamma_\infty}\omega_\gamma^{a_\gamma b_\gamma}\right\}\sin(qx)}{2\pi i \rho C_\infty\left\{\omega\left[\omega_C^{a_C}+(i\omega)^{a_C}\right]^{b_C}+\frac{\Delta C}{C_\infty}\omega\omega_C^{a_Cb_C}-i\alpha_\infty q^2\left[\omega_C^{a_C}+(i\omega)^{a_C}\right]^{b_C}\right\}\left(\omega-\omega_3\right)\left(\omega-\omega_4\right)\left[\omega_\gamma^{a_\gamma}+(i\omega)^{a_\gamma}\right]^{b_\gamma}}.
\label{u_omega_substituted_HN}
\end{equation}
The ISS signal can then be obtained in analogy with \ref{SubSec:Freq_dep_u}.

\end{widetext}

\section{Comparison with stretched exponential model \label{Sec:comparison_with_liter}}

\begin{figure}
\begin{center}
\includegraphics[width=0.38\textwidth]{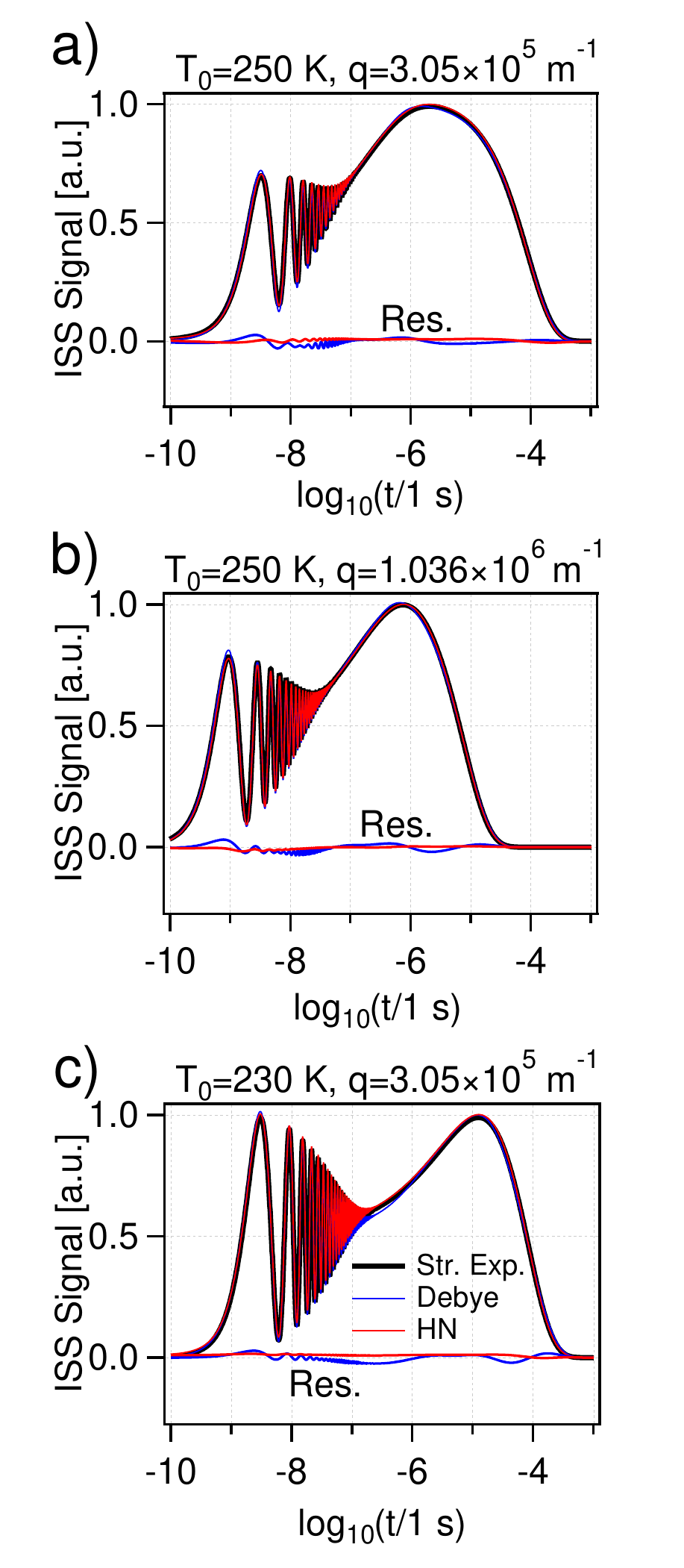} 
\caption{Plot of the time dependence of the ISS signal obtained with the SEM (black curves), most squares fit of this signal with the Debye model (blue curves) and the HN model (red curves), for different temperature-wavenumber combinations, based on material parameters of glycerol reported in \cite{paolucci2000impulsive} and listed in Table \ref{table_Param}. The small fitting residues (fitting curve minus SEM curve) indicate that the Debye based model is adequate.  Thanks to the two additional model parameters, the HN model is fitting even better.
In each panel, all the curves are normalized to the maximum of the SEM ISS signal.}
\label{ISS_strVsHN}
\end{center}
\end{figure}

A semi-empirical model (SEM) describing ISS signals in glassformers has been introduced along with the first experimental reports on ISS signals in salol \cite{yang1995t,yang1995impulsive1,yang1995impulsive2} and has been successfully used to fit ISS data in glycerol \cite{paolucci2000impulsive}.
The SEM expression describing the ISS signal is \footnote{
Since in this work we study the ISS signal detected with a heterodyne experimental setup, the right-hand-side of Eq. \ref{Str_exp_expr} is not squared. This choice is at variance with respect to Ref. \cite{paolucci2000impulsive}, which is based on homodyne detection.
}:
$$I(t)=(A+B)\exp\left(-\Gamma_Ht\right)+$$
\begin{equation}
-A\exp\left(-\Gamma_At\right)\cos\left(\omega_At\right)-B\exp\left[-\left(\Gamma_Rt\right)^\beta\right],
\label{Str_exp_expr}
\end{equation} 
where the first term is associated to the thermal diffusion ($\Gamma_H$ being the thermal decay rate), while the second one corresponding to the acoustics ($\omega_A$ being the acoustic oscillation frequency and $\Gamma_A$ the acoustic damping rate).
The use of a stretched exponential  - also known as Kohlrausch-Williams-Watts (KWW) - term was inspired by other response functions in the physics of supercooled liquids, and was aimed at coping with the empirical observation that the initial thermal expansion cannot be fitted by a simple exponential.
$\Gamma_R$ is the structural relaxation rate and $0<\beta\leq 1$ the stretching exponent.
The coefficients $A$ and $B$ account for the weights of each term contributing to the total ISS signal.\\
The SEM has proved useful to describe and fit the ISS signal measured on supercooled glycerol, as described by Paolucci et al. \cite{paolucci2000impulsive}.
The fitting parameters of interesting temperature-wavenumber combinations treated by Paolucci et al. are recalled in Table \ref{table_Param} and were used to simulate respective theoretical ISS signals, shown in Fig. \ref{ISS_strVsHN} (black curves).
\begin{table}[h]
\begin{center}
\begin{tabular}{|c|c|c|c|c|}\hline 
Case  &\#1 & \#2&\#3&\#4 \\ \hline
$T_0$ (K) & 250 & 250 & 230 & 230 \\ \hline 
$q$ (m$^{-1}$) & 3.05$\times 10^5$ & 1.036$\times 10^6$ & 3.05$\times 10^5$ & 1.036$\times 10^6$ \\ \hline 
$\Gamma_H$ (s$^{-1}$)\footnote{
The thermal decay rate was estimated as $\Gamma_H=\kappa_T q^2/(\rho\ C)$, where $\kappa_T=0.28\  \um{W/m\ K}$, $\rho=1260\ \um{kg/m^3}$ \cite{gupta2012scope} and $C=1800$ J/kg K (for T$_0=250$ K) or $C=1500$ J/kg K (for T$_0$=230 K), in agreement with Refs. \cite{bentefour2003broadband,bentefour2004thermal}.
} & 1.14$\times 10^4$ & 1.33$\times 10^5$ & 1.38$\times 10^4$ & 1.59$\times 10^5$ \\ \hline 
$\Gamma_A$ [s$^{-1}$]\footnote{
The acoustic damping rate was taken from Fig. 3 of Ref. \cite{paolucci2000impulsive}.
} & 3.5$\times 10^7$ & 8.5$\times 10^7$ & 2.0$\times 10^6$ & 2.5$\times 10^6$ \\ \hline
$\omega_A$ (Grad/s)\footnote{
The acoustic oscillation frequency was obtained as the product between $q$ and the speed of sound reported in Fig. 3 of Ref. \cite{paolucci2000impulsive}.
} & 0.98 & 3.32 & 1.03 & 3.49 \\ \hline 
$\beta$\footnote{
The stretch exponent was taken from Fig. 5 of Ref. \cite{paolucci2000impulsive}.
} & 0.6 & 0.6  & 0.6  & 0.6 \\ \hline
$\Gamma_R$ (s$^{-1}$)\footnote{
The structural relaxation rate was obtained as $\Gamma_R=\Gamma({1/ \beta})/\left(\langle\tau\rangle\beta \right)$, where $\Gamma$ is the Gamma function, and $\langle\tau\rangle$ was taken from Fig. 4 of Ref. \cite{paolucci2000impulsive}.
} & 5.5$\times 10^6$ & 5.5$\times 10^6$ &1.1$\times 10^5$ & 1.1$\times 10^5$ \\ \hline
$B/A$\footnote{
The ratio between the coefficients $B$ and $A$ was calculated as $B/A=f/(1-f)$, where $f=0.67$ is the Debye-Waller factor taken from Fig. 7 of Ref. \cite{paolucci2000impulsive}.
} & 2.03 & 2.03  & 2.03  & 2.03 \\ \hline
\end{tabular}
\end{center}
\caption{Material parameters reported by Paolucci et al. based on SEM fits of ISS signals in supercooled glycerol.
The parameters reported in case \#4 imply a SEM model ISS signal with the unphysical negative tail, as reported in Fig. \ref{Case4_ISS_signal}. This unphysical behavior is not present in the other cases.}
\label{table_Param}
\end{table}
Three cases are considered in Fig. \ref{ISS_strVsHN}: in panel a and b the ISS of glycerol is reported for the two grating wave numbers $q=3.05\times10^5\ \um{m^{-1}}$ and $q=1.036\times10^6\ \um{m^{-1}}$ respectively, at the same temperature $T_0=250$ K (i.e. cases \#1 and \#2 in Table \ref{table_Param}, respectively).
Panel c reports the ISS signal for a lower temperature $T_0=230$ K and for the shortest grating wave vector available, i.e. $q=3.05\times10^5\ \um{m^{-1}}$ (case \#3 in Table \ref{table_Param}). For each panel, the value of the coefficient $A$ was chosen in order to have the maximum of the ISS signal normalized to 1.\\
In order to verify if the Debye and HN models developed in the previous section are able to reproduce the SEM based ISS signal, we have fitted the black curves in Fig. \ref{ISS_strVsHN} with the respective expressions (blue curves: Debye, red curves: HN).
The fitting was carried out by implementing a most-squares fitting (MSF) protocol \cite{jackson1976most,salenbien2011laser} to search for the minimum of the cost function, defined as the sum of the squared residuals (SSR).
MSF is advantageous over the commonly used least-squares fitting (LSF) as it is able to take into account the possible co-variance of the multiple fitting variables, namely, different combinations of fitting parameters yielding a statistically indistinguishable cost function value SSR (local minima).
Both models fit very well, with the residues of the HN model being the smallest, thanks to the additional two fitting parameters.
\begin{table*}
\begin{center}
\begin{tabular}{|c|c|c|c|}\hline
Case	&	\#1	&	\#2	&	\#3	\\ \hline
$C_{\infty}$ Debye (J/kg K)	&	452	&	460	&	360	\\ \hline
$C_{\infty}$ HN (J/kg K)	& $503\in[0,6\times 10^5]$ & $503\in[0,6\times 10^5]$ & $390\in [388,400]$\\ \hline
$C_0$ Debye (J/kg K)	&	1837	&	1882	&	1645	\\ \hline
$C_0$ HN (J/kg K)	&	$1777\in[1775,1798]$	&	$1777\in[1775,1798]$	&	$1345\in[1340,1351]$	\\ \hline
$\Delta \gamma/\gamma_\infty$ Debye	&	10	&	10	&	10	\\ \hline
$\Delta \gamma/\gamma_\infty$ HN	& $9.4\in[9.3,9\times 10^3]$	&	$9.4\in[9.3,9\times 10^3]$	&	$9.8\in[9.7,10]$	\\ \hline
$\omega_C$ (Mrad/s) Debye	&	5.1	&	6.6	&	0.2	\\ \hline
$\omega_C$ (Mrad/s) HN	&$3.5\in[0,4\times10^3]$&$3.5\in[0,4\times10^3]$&$1.5\in[1.1,500]$\\ \hline
$\omega_\gamma$ (krad/s) Debye	&	3190	&	4080	&	150	\\ \hline
$\omega_\gamma$ (krad/s) HN	&	$2260\in[2230,2320] $	&	$2260\in[2230,2320] $&	$38\in[0,39]$	\\ \hline
$a_c$ HN	&	$0.89\in[0.06,0.9]$	&	$0.89\in[0.06,0.9]$	&	$0.544\in[0.543,0.554]$	\\ \hline
$b_c$ HN	&	$0.52\in[0.02, 0.55]$	&	$0.52\in[0.02, 0.55]$	&	$0.78\in[0.75,1]$	\\ \hline
$a_\gamma$ HN	&	$0.9\in[0,1]$&$0.9\in[0,1]$	&$0.7\in[0,1]$	\\ \hline
$b_\gamma$ HN	&	$0.68\in[0.67,1]$	&	$0.68\in[0.67,1]$	&	$0.5\in[0.1,1]$	\\ \hline
\end{tabular}
\end{center}
\caption{Fitting parameter values obtained by fitting the SEM curves reported in Fig. \ref{ISS_strVsHN} with Debye and HN model based expressions for the ISS signal. For the HN model, we report also confidence interval next to the best fit parameter.}
\label{table_fitting_coeff}
\end{table*}

The obtained fitting parameters are summarized in Table \ref{table_fitting_coeff}. For the HN model, for each parameter, the fitting error was determined by most squares analysis of the cost function in the multidimensional space of fitting parameters and thus includes the effect of covariance with other fitting parameters \cite{jackson1976most,salenbien2011laser}.
Instead of displaying the parameter $\Delta C$ obtained from the fit, in Table \ref{table_fitting_coeff} we have reported the zero-frequency heat capacity $C_0=C_\infty+\Delta C$. The latter definition was retrieved from Eq. \ref{C_omega} in the limit $\omega\rightarrow0$.

The fits with the HN model expression have been performed simultaneously on cases $\#1$ and $\#2$, where the glycerol equilibrium temperature is the same, but the grating wave numbers are different. In ISS signals, besides a change of acoustic frequency and damping, a difference in wave number results (when viewing the signal on a logarithmic time scale) in a different "onset" time ($1/q^2\alpha$) of the thermal diffusion driven washing out of the thermal (expansion) grating. Once ongoing, this exponential decay dominates the signal behavior and masks the influence of the parameters that determine the onset of the relaxation of the temperature to heat and thermal expansion to temperature response. Simultaneously fitting signals at two wavenumbers, and thus considering signals containing two mixing ratios of the respective influences, limits the possibilities for covariant influences of fitting parameter values on the signals, and thus leads to smaller uncertainties.
   
Interestingly, the uncertainty on $C_0$ is always small. This can be explained by the strong influence of the low frequency/late time limit of the heat capacity on the signal, via the always significant thermal diffusion related exponential decay tail of the signal. For the chosen signals, this tail occurs later than the relaxation times, and it is thus not affected by possible degeneracy of the effect of $C_0$ on the signal with relaxation influenced thermal expansion parameters. The influence of $C_\infty$ on the initial part of the signal goes along with influence of the thermal expansion and the acoustic wave related parameters. This leads to a larger fitting covariance.   

\begin{figure}
\begin{center}
\includegraphics[width=0.4\textwidth]{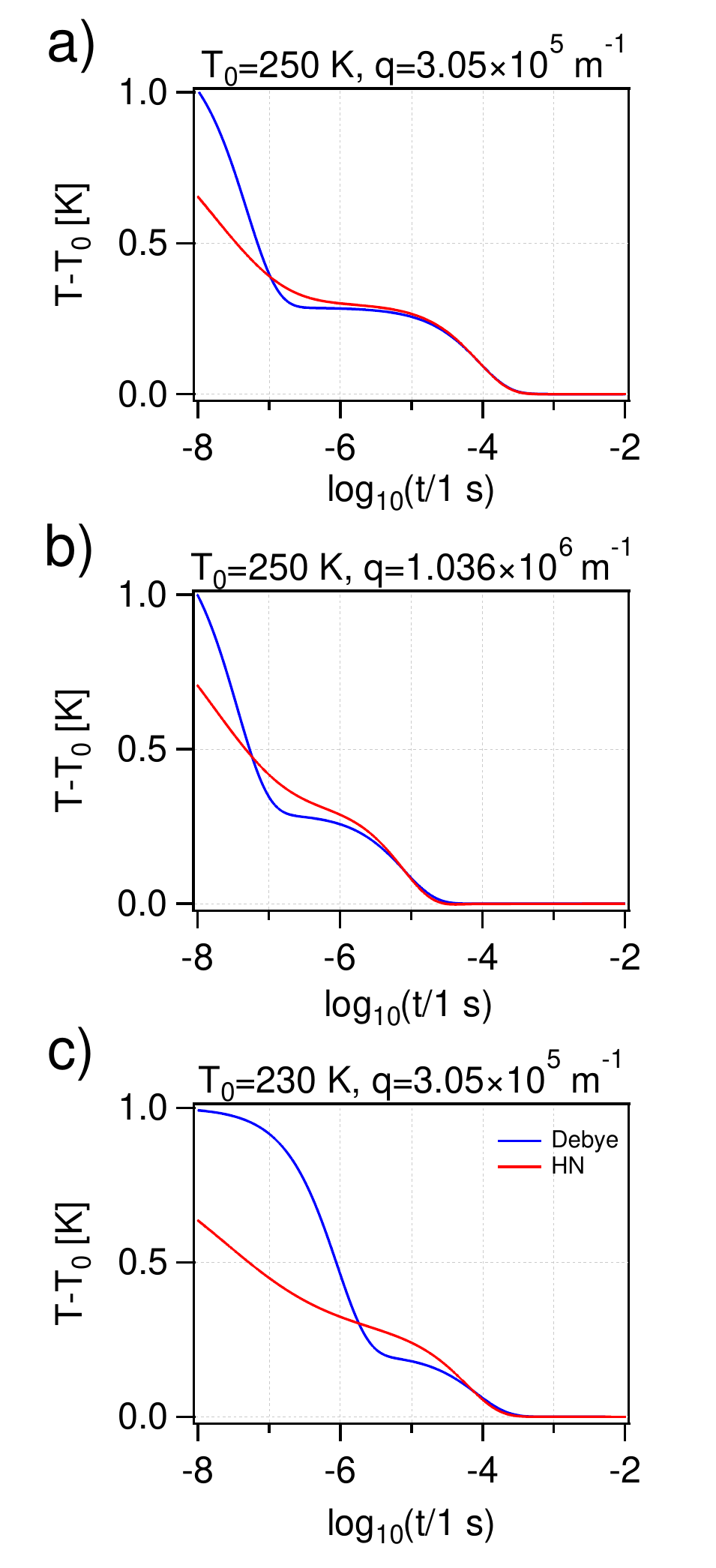} 
\caption{Time evolution of the temperature obtained for the case of Debye and HN models (blue and red curves, respectively), simulated for different temperature-wavenumber combinations.
In each panel, all the curves are normalized to the value of the Debye model temperature at the shortest displayed time\protect\footnote{
Eq.s \ref{T_t_expression_final} and \ref{temperature_in_freq_C_omega_HN} depend on the spatial coordinate.
The curves in Fig. \ref{temperature_strVsHN} were evaluated at the same spatial coordinate.
The particular choice of the latter is irrelevant thanks to the proposed normalization.}.}
\label{temperature_strVsHN}
\end{center}
\end{figure}

\begin{figure}[h]
\begin{center}
\includegraphics[width=0.41\textwidth]{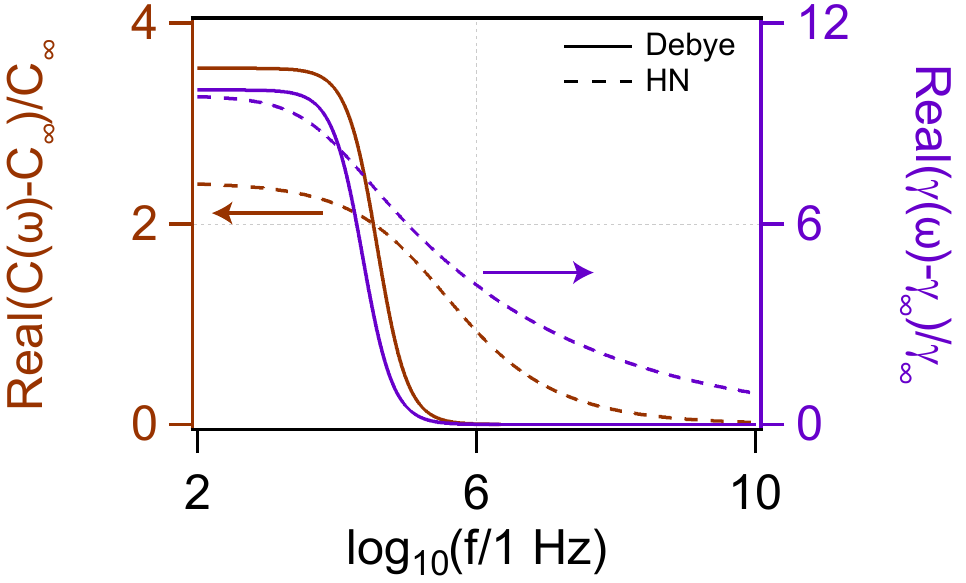} 
\caption{Real part of $\left(C(\omega)-C_\infty\right)/C_\infty$ (brown lines, left axis) and of $\left(\gamma(\omega)-\gamma_\infty\right)/\gamma_\infty$ (purple lines, right axis), as a function of frequency (horizontal axis, log scale). 
The full and dashed lines refer to the Debye and HN model, respectively.
These curves were calculated for the case $T_0=230$ K and $q=3.05\times 10^5\ \um{m^{-1}}$ (case $\#3$).
}
\label{C(omega)_Gamma(omega)}
\end{center}
\end{figure}

The limited effect of $C_\infty$ on the ISS signal can be further understood by looking at the calculated temperature evolution after impulsive illumination, as depicted in Fig. \ref{temperature_strVsHN} for the Debye and HN models (blue and red curves, respectively).
The temperature evolution for the Debye model was obtained by evaluating Eq. \ref{T_t_expression_final} upon insertion of the Debye model parameters listed in Table \ref{temperature_strVsHN}.
For the calculation based on the HN model, the thermal parameters were first inserted into Eq. \ref{temperature_in_freq_C_omega_HN} to calculate the HN temperature in frequency domain.
The latter was then Fourier transformed numerically to time domain, obtaining the red curves in Fig. \ref{temperature_strVsHN}.
For long times the Debye and HN model based temperature evolutions match.  However, despite the fact that the Debye and HN model yield a very similar ISS signal (as shown in Fig. \ref{ISS_strVsHN}), the corresponding fitting parameters reported in Table \ref{temperature_strVsHN} imply a very different temperature profile at short times.
This indicates a very strong degeneracy between the early dynamics of the temperature, the thermal expansion and the acoustic wave generation.\\
From another point of view, one may wonder why two different temperature profiles can give rise to the same ISS signal.
This can also be seen by looking into the math: substituting the expression for the temperature into the source term of the Eq. \ref{Expansion_isotropic_FT_compact} for the displacement, we see that the source term is proportional to $\gamma/\left[\rho C\left(\omega-i\alpha q^2\right)\right]$ (as reported in Eq. \ref{def_Z}). 
Hence, at high frequencies the source term is proportional to $\gamma/C$.
Fig. \ref{C(omega)_Gamma(omega)} shows that the real part of $C(\omega)$ and $\gamma(\omega)$ for cases $\#3$ follow a very similar trend, both for the Debye and HN model.
We verified that this is also the case for the imaginary part of these quantities. 
Hence, even if the time dependences of $C$ and $\gamma$ are quite different between the Debye and HN scenario, their ratio, and thus the corresponding ISS signal is similar.\\
A similar observation has been made for cases $\#1$ and $\#2$.\\
In conclusion, at short times the parameters $C$ and $\gamma$ are degenerate and their individual values cannot be reliably extracted from fitting.
Conversely, at low frequency the source term is no longer simply proportional to the ratio $\gamma/C$, hence the degeneracy is lifted and at long times the heat capacity and thermal expansion coefficient can be disentangled precisely by the the fitting procedure. A prospective scenario for lifting the large degeneracy between $C_\infty$ and $\gamma_\infty$ could be to further increase the wavenumber so that the thermal diffusion tail occures before the heat capacity relaxation time. In that scenario, the shape of the thermal diffusion tail is dominated by the decay time, which gives direct information on $C_\infty$, with little influence of the other parameters.  \\

\begin{figure}[h]
\begin{center}
\includegraphics[width=0.36\textwidth]{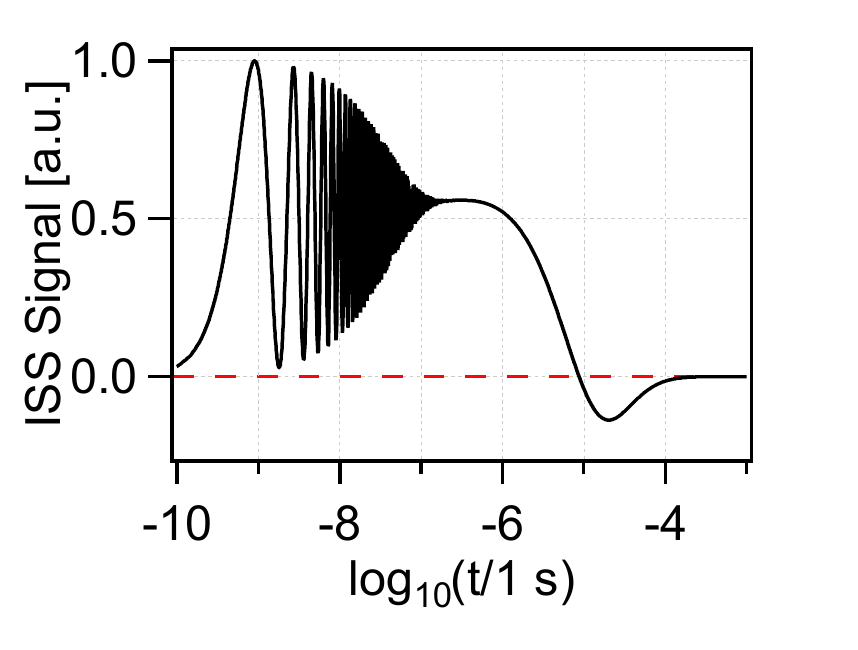} 
\caption{Plot of the ISS signal obtained with SEM (black curves), for the case $T_0=230$ K and $q=1.036\times 10^6\ \um{m^{-1}}$ (case $\#4$).
The curve is normalized to 1 at the maximum. The graph shows that the ISS signal goes below zero (dashed red line).}
\label{Case4_ISS_signal}
\end{center}
\end{figure}

It is worth to note that the ISS signal obtained for case $\#4$, which is representative for a rather low temperature (long relaxation times) and a rather long grating spacing (long thermal diffusion time) in Fig.  \ref{Case4_ISS_signal} reveals a temporal span in which the ISS signal, calculated according to the SEM model, is negative.

This unphysical result, which is a consequence of a particular mix between positive and negative terms in Eq. \ref{Str_exp_expr}, prevents an adequate comparison with the here presented models, both in the Debye and HN scenario.  
For the sake of curiosity, we have evaluated the conditions for which the SEM based ISS signal goes negative.
For this evaluation, we have simplified Eq. \ref{Str_exp_expr} by neglecting the acoustic term with respect to the thermal diffusion, because (i) the amplitude of the former ($A$) is smaller than the one of the latter ($A+B$) and (ii) the former decays faster than the latter ($\Gamma_A$ being much larger than $\Gamma_H$).
Upon this simplification, the ISS signal in the SEM model becomes:
\begin{equation}
I(t)=(A+B)\exp\left(-\Gamma_Ht\right)-B\exp\left[-\left(\Gamma_Rt\right)^\beta\right].
\label{Str_exp_expr_simplified}
\end{equation}
By performing some algebric calculations, one obtains that $I(t)$ is positive for time instants satisfying the following inequality:
\begin{equation}
t^\beta\geq\left(\Gamma_H\over \Gamma_R^\beta\right)t-{1\over \Gamma_R^\beta}\ln\left(1+\frac{A}{B}\right).
\label{inequality_I_bigger_0}
\end{equation}
It is evident that for the limit $t\rightarrow0$, Eq. \ref{inequality_I_bigger_0} is satisfied, while in the limit $t\rightarrow +\infty$, Eq. \ref{inequality_I_bigger_0} is violated.
Hence, there must be at least one time where $I(t)$ changes sign.
We call $t^*$ the earliest positive time satisfying $I(t)=0$.\\
Furthermore, since $\beta<1$ for the cases reported in Table \ref{table_Param}, the slope of the left-hand side of Eq. \ref{inequality_I_bigger_0} decreases for increasing time, while the slope of the right-hand side is constant.
Therefore, for times later than $t^*$, the right-hand-side grows faster than the left-hand-side and hence, there are not other time instants satisfying $I(t)=0$.
Summarizing, for $0\leq t\leq t^*$, the ISS signal is positive, while for $t>t^*$ the ISS signal is negative.\\
The derivation of an analytic expression for $t^*$ is challenging.
However, we can have some insight by speculating on the time instant:
\begin{equation}
t^*_{low}=\left(\Gamma_H\over \Gamma_R^\beta\right)^{1\over(\beta-1)},
\label{t_star_low}
\end{equation}
which is a lower bound for $t^*$ \footnote{
$t^*_{low}$ is the time instant solving Eq. \ref{inequality_I_bigger_0} without the last term, the latter reading:
$$t^\beta=\left(\Gamma_H\over \Gamma_R^\beta\right)t.$$
Hence, we have 
$$\left(t^*_{low}\right)^\beta=\left(\Gamma_H\over \Gamma_R^\beta\right)t^*_{low}>\left(\Gamma_H\over \Gamma_R^\beta\right)t^*_{low}-{1\over \Gamma_R^\beta}\ln\left(1+\frac{A}{B}\right).$$
The latter relation states that $t^*_{low}>t^*$ or, in other words, $t^*_{low}$ is a lower bound for $t^*$}.\\
In Table \ref{table_t_star_low} we have evaluated $t^*_{low}$ for the four cases considered in the current section.
\begin{table}[h]
\begin{center}
\begin{tabular}{|c|c|c|c|c|}\hline 
Case  &\#1 & \#2&\#3&\#4 \\ \hline
$t^*_{low}$ (s) & 0.93 & $2.0\times 10^{-3}$ &  $1.6\times 10^{-3}$ &  $3.6\times 10^{-6}$ \\ \hline 
\end{tabular}
\end{center}
\caption{Coefficients reported by Paolucci et al. concerning the fitting of the supercooled glycerol ISS signals with SEM model.}
\label{table_t_star_low}
\end{table}
For cases $\#1$ to $\#3$, the lower bound for $t^*$ (and, hence, $t^*$ itself) occurs on only after several milliseconds.
Therefore, the negative ISS signal is not visible in Fig. \ref{ISS_strVsHN}: it occurs on late times that are not shown in the figure.
For case $\#4$ $t^*_{low}\sim 3.6\ \um{\mu s}$, paving the way to the detection of a negative ISS signal for times $t>t^*\sim 10\ \um{\mu s}$. This is confirmed in Fig. \ref{Case4_ISS_signal}.\\

\section{Debye model vs two-temperature model\label{Sec:two_temp_model}}
\subsection{Frequency dependent heat capacity \label{Subsec:two_temp_model_HC}}
As mentioned earlier, part of the energy that is optically supplied to a relaxing material is channeled, around the relaxation time, to a change of configurational energy that goes along with a structural rearrangement of the amorphous network.
In this section, we describe the network's energy distribution in terms of a temperature $T_N$, and we assume that the configurational energy reservoir is in thermal contact with a kinetic (mainy vibrational) energy reservoir (KER), with physically measurable temperature T.
The energy flux between the two reservoirs is quantified as $G(T-T_N)$, where $G$ (W/(m$^3$ K)) is a (positive) coupling constant; this term indicates that when $T>T_N$, then energy flows from the KER to the network.
The capability of storing and transferring energy within the network are formally described by the network's heat capacity $C_N$, thermal conductivity $k_{T,N}$ and density $\rho_N$.\\
Hence, in this approach, the energy exchange between the KER (vibrational energy) and the network (configurational energy) can be described by a two-temperature model (TTM) \cite{caddeo2017thermal}, yielding the following equations for the KER's temperature $T$ and the network's temperature $T_N$:\\
\begin{equation}
\left\{\begin{array}{l}
\displaystyle{\rho C \frac{\partial T}{\partial t}=\kappa_T\frac{\partial^2 T}{\partial x^2}+Q(x,t)-G(T-T_N)},\\
\\
\displaystyle{\rho_N C_N \frac{\partial T_N}{\partial t}=k_{T,N}\frac{\partial^2 T_N}{\partial x^2}+G(T-T_N)},\\
\end{array}
\right.
\label{two_temperatures_model}
\end{equation}
where $\rho$, $C$ and $\kappa_T$ are the KER's density, heat capacity and thermal conductivity, respectively.
The source term $Q$ (W/m$^3$) enters only in the equation for the KER temperature, indicating that the optical excitation delivers energy directly to the KER. The network modifications follow the dynamics occurring in the KER, hence they are driven by variations of $T$.\\
For the sake of simplicity, we assume that local network reconfigurations only depend on the local energy exchange with the KER, and we neglect possible configurational energy flow within the network.
This yields a very low value for the network's thermal conductivity, allowing to drop the term $k_{T,N}{\partial^2 T_N}/\partial x^2$ in System \ref{two_temperatures_model}.\\
The first equation of System \ref{two_temperatures_model} can be reformulated as:
\begin{equation}
T_N=T+\frac{\rho C}{G} \frac{\partial T}{\partial t}-\frac{\kappa_T}{G}\frac{\partial^2 T}{\partial x^2}-{1\over G}Q(x,t).
\label{first_TTM_rewritten}
\end{equation}
Substituting the latter expression into the second equation of System \ref{two_temperatures_model}, we get to the following differential equation:
$$\frac{\rho_N C_N}{G}\left(\rho C \frac{\partial^2 T}{\partial t^2}-\frac{\partial Q}{\partial t}+G\frac{\partial T}{\partial t}-\kappa_T\frac{\partial^3 T}{\partial t \partial x^2} \right)=$$
\begin{equation}
=-\rho C\frac{\partial T}{\partial t}+Q(x,t)+\kappa_T\frac{\partial^2 T}{\partial x^2}.
\label{decoupled_for_T}
\end{equation}
Applying a Fourier transform to Eq. \ref{decoupled_for_T} we get to:
$$\frac{\rho_N C_N}{G}\left[-\omega^2\rho C \tilde{T}-i\omega \tilde{Q}(x,\omega)+i\omega G\tilde{T}-i\omega \kappa_T\frac{\partial^2 \tilde{T}}{ \partial x^2} \right]=$$
\begin{equation}
=-i\omega\rho C\tilde{T}+\tilde{Q}(x,\omega)+\kappa_T\frac{\partial^2 \tilde{T}}{\partial x^2},
\label{decoupled_for_T_freq}
\end{equation}
which can be rearranged as:
$$\kappa_T\left(1+i\frac{\rho_N C_N}{G}\omega\right)\frac{\partial^2 \tilde{T}}{\partial x^2}+$$
$$+\left[\frac{\rho_N C_N \rho C}{G}\omega^2-i\omega\left(\rho_N C_N +\rho C\right)\right]\tilde{T}+$$
\begin{equation}
+\left(1+i\omega\frac{\rho_N C_N}{G} \right) \tilde{Q}(x,\omega)=0.
\label{decoupled_for_T_freq_2}
\end{equation}
Dividing by $\kappa_T(1+i\omega \rho_N C_N/G)$ and rewriting the second term we have:
\begin{equation}
\frac{\partial^2 \tilde{T}}{\partial x^2}-i\omega\frac{\rho}{\kappa_T}\left[C+\frac{{\rho_N \over \rho} C_N}{1+i\frac{\rho_N C_N}{G}\omega}\right]\tilde{T}=-{1\over \kappa_T}\tilde{Q}(x,\omega).
\label{decoupled_for_T_freq_4}
\end{equation}
The latter equation can be remapped into Eq. \ref{diffusion_equation_in_freq_C_omega} upon substitutions $C\rightarrow C_\infty$, $\rho_N C_N/\rho\rightarrow \Delta C$ and $G/(\rho_N C_N)\rightarrow \omega_C$.
Hence, it can be stated that the frequency dependent heat capacity in terms of the Debye model (as reported in Eq. \ref{C_omega}) is equivalent to the TTM.\\

\subsection{Frequency dependent thermal expansion in the frame of the two-temperature model
\label{Sec:two_temp_model_therm_exp}}
Along with the process of taking up potential energy, the network is undergoing structural changes, and it can thus change its volume.
Hence, it can contributute to the system's thermal expansion and add up to the thermal strain that is related to the increase of vibrational energy (which is connected to the anharmonicity of the intermolecular potential minima).  
To the best of our knowledge, no models describing the latter point are reported in literature.
Nevertheless, it seems reasonable to assume that the thermal strain produced both by the KER and the network depends on the history of the network's temperature.
Assuming an isotropic material, the total thermal strain can hence be written as $\left[\gamma \Delta T\circledast \varphi(t)+\gamma_{N} \Delta T_N\circledast \varphi_N(t)\right]I_d$, where $I_d$ is the identity matrix.
Indeed, the two temperatures have been convolved with memory functions $\varphi(t)$ and $\varphi_N$ describing the thermal history of the KER and network respectively.\\
After substituting $\gamma_M \Delta T(x,t)$ with $\left[\gamma \Delta T\circledast \varphi(t)+\gamma_{N} \Delta T_N\circledast \varphi_N(t)\right]I_d$ into Eq \ref{definition_of_strain_C}, we can repeat analogously the derivation presented in Section \ref{Sec:displacement}.\\
In this way, we reach to the following equation for the displacement ($\xi(\omega)$ is defined in Eq. \ref{def_xi}):

$$\frac{\partial^2 \tilde{u}_x}{\partial x^2}+\frac{\omega^2}{c^2(\omega)} \tilde{u}_x=$$
$$=\xi(\omega)\frac{\partial }{ \partial x}\left(\gamma\tilde{T}(x,\omega)\tilde{\varphi}(\omega)+\gamma_N\tilde{T}_N(x,\omega)\tilde{\varphi}_N(\omega)\right)=$$

\begin{equation}
=\xi(\omega)\left(\gamma\tilde{\varphi}(\omega)+\frac{\gamma_N\tilde{\varphi}_N(\omega)}{1+i\frac{\rho_N C_N}{G}\omega}\right)\frac{\partial \tilde{T}}{ \partial x}.
\label{Expansion_isotropic_FT_compact_TN}
\end{equation}
The last step involved the substitution $\tilde{T}_N=\tilde{T}(x,\omega)/\left(1+i\frac{\rho_N C_N}{G}\omega\right)$ (see Appendix \ref{app:Derivation of the} for the proof of the latter expression).\\
We suppose that the network contribution to the thermal strain contains an instantaneous term and a second term accounting for the network's thermal history. 
These considerations are well reproduced by a memory function of the type:
\begin{equation}
\varphi_N(t)=2\pi\delta(t)+2\pi\omega_C\left(1-\chi_\gamma\right)\exp(-\omega_\gamma t)\theta(t)
\label{def_of_varphi}
\end{equation}
which corresponds to the following expression for the network's contribution to the thermal strain:
$$\gamma_{N} \Delta T_N\circledast \varphi_N(t)=2\pi \gamma_{N}\Delta T_N(x,t)+$$

\begin{equation}
+2\pi\omega_C\left(1-\chi_\gamma\right)\int_{-\infty}^{t}\exp\left[-\omega_\gamma(t-\tau)\right]\Delta T_N(\tau)d\tau.
\label{equivalence_therm_strain_varphi}
\end{equation}
With the latter definitions, the past thermal events in the network are exponentially less and less important with increasing time in the past.
The temporal cutoff for the exponential is the inverse of a frequency $\omega_\gamma$, which can be written in terms of the frequency $\omega_C=G/(\rho_NC_N)$ (already introduced in Subsection \ref{Subsec:two_temp_model_HC}) as:
\begin{equation}
\omega_\gamma=\chi_\gamma \omega_C=\chi_\gamma\frac{G}{\rho_NC_N}.
\label{chi_gamma}
\end{equation}
$\chi_\gamma$ quantifies to what extent the thermal expansion relaxation is slower than the heat capacity relaxation.\\
The Fourier transform of the memory function reads \footnote{The Fourier transform of $2\pi\delta(t)$ is $1$. Furthermore, the Fourier transform of $e^{-A t}\theta(t)$, with $A$ real and strictly positive, is $(2\pi)^{-1}(i\omega+A)^{-1}$.}:
\begin{equation}
\tilde{\varphi}_N(\omega)=1+\frac{\omega_C\left(1-\chi_\gamma\right)}{i\omega+\omega_\gamma}={1\over\chi_\gamma}\frac{1+i{\omega\over \omega_C}}{1+i{\omega\over \omega_\gamma}}.
\label{phi_omega}
\end{equation}
By introducing Eq. \ref{phi_omega} into Eq. \ref{Expansion_isotropic_FT_compact_TN} we get to:
\begin{equation}
\frac{\partial^2 \tilde{u}_x}{\partial x^2}+\frac{\omega^2}{c^2(\omega)} \tilde{u}_x=\xi(\omega)\left(\gamma\tilde{\varphi}(\omega)+\frac{\gamma_N/\chi_\gamma}{1+i{\omega\over\omega_\gamma}}\right)\frac{\partial \tilde{T}}{ \partial x}.
\label{Expansion_isotropic_FT_compact_semi_final}
\end{equation}
We should also provide an expression for the KER's memory function.
If we assume that only the instantaneous value of $T$ is important for the evaluation of the thermal strain, i.e. $\varphi(t)=2\pi\delta(t)$, then $\tilde{\varphi}(\omega)$ becomes identically 1, yielding:
\begin{equation}
\frac{\partial^2 \tilde{u}_x}{\partial x^2}+\frac{\omega^2}{c^2(\omega)} \tilde{u}_x=\xi(\omega)\left(\gamma+\frac{\gamma_N/\chi_\gamma}{1+i{\omega\over\omega_\gamma}}\right)\frac{\partial \tilde{T}}{ \partial x}.
\label{Expansion_isotropic_FT_compact_final}
\end{equation}

By performing the substitution $\gamma\rightarrow \gamma_\infty$ and $\gamma_N/\chi_\gamma\rightarrow \Delta \gamma$, the latter equation can be mapped on Eq. \ref{Expansion_isotropic_FT_compact_gamma_omega}.
Again, the choice of the Debye model for $\gamma$ (relation \ref{gamma_omega}) can be justified in terms of the TTM.\\
Equivalently, the choice of the thermal expansion in the frame of Debye model implies that the thermal strain is related instantaneously to the KER's temperature. On the other hand, the thermal history of the network has to be accounted for to estimate the thermal strain. The Debye model implies that the memory function for the network is described by Eq. \ref{def_of_varphi}.\\
Analogously, the thermal expansion ruled by the HN model can be justified considering the thermal history of the KER and of the network.
However, the complexity of the HN model, yielding also non-integer exponentials, prevents the possibility of having a simple and general expression for the memory functions.\\

\section{Experimental results and discussion}
\label{experiment}
We have heretofore developed the generalized physical model addressing the ISS response of glass-forming liquids subject to ultrafast photothermal excitation.
\begin{figure}[h]
\begin{center}
\includegraphics[width=0.5\textwidth]{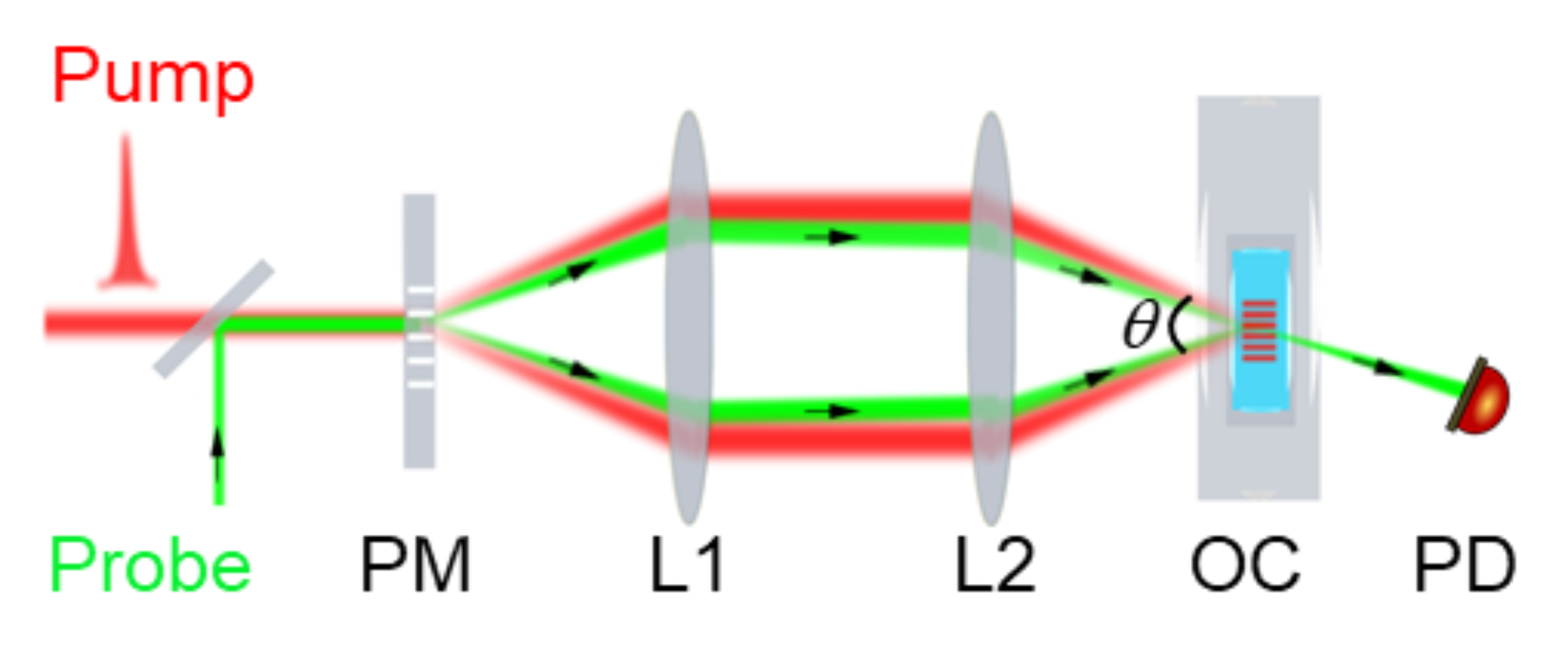} 
\caption{Scheme of the experimental setup based on the heterodyne-detected transient grating technique. A spatially periodical laser pattern from a pulsed pump laser (red) is formed on the sample to create thermoelastic transients, the latter detected by a coaxially aligned probe laser (green and arrows).
In the scheme we sketch the phase mask (PM), the lenses (L1 and L2), the optical cryostat (OC) and the photodetector (PD).} 
\label{exp_setup}
\end{center}
\end{figure}
Now, an experimental study of the ISS response of glycerol ($>99.0 \%$ purity) under supercooling is presented. An ultrafast heterodyne-detected transient grating (HD-TG) setup is used for the experiment.
Fig. \ref{exp_setup} shows the scheme of the setup, in which a ps pump laser pulse at 1064 nm (shown in red) is diffracted by a transmission phase mask (PM) into two 1st diffraction orders, namely $\pm1$ orders.
The two are then recombined via a two-lens (4\textit{f}) imaging system into the bulk of the sample. The sample is accommodated in a liquid nitrogen optical cryostat (OC) to allow temperature control over it. The light interference forms a spatially periodical light pattern and creates a transient local density grating (thermoelastic transients), at a wavelength identical to the spacing of the light grating, $d$.

For a given light wavelength light $\lambda$, one can tune the spacing of the excitation pattern by varying the intersecting angle of the two beams $\theta$, namely via $d=\lambda/(2 n) \sin{(\theta/2)}$ with $n$ the optical refractive index of the sample medium. 
In this setup, the $\theta$-tuning is implemented by translating a phase mask (PM) array containing multiple PMs of varying period \cite{verstraeten2015determination}.
Alternatively, one can also rotate the PM to realize the $\theta$-tuning \cite{vega2015laser}.
The detection of ISS takes advantage of the optical heterodyne scheme \cite{maznev1998optical}, in which the probe beam from a continuous wave (CW) laser at wavelength 532 nm (shown in green with black arrows in Fig. \ref{exp_setup}), is aligned to be coaxial with the pump beam.
Both beams are sent to the PM and diffracted into excitation and probe/reference beam pairs. This heterodyne scheme has been widely used in the field for studying optical transparent or weakly absorbing liquids \cite{brodard2005application,taschin2008time,glorieux2002thermal} owing to its high sensitivity. More detailed description of the setup can be found in Ref. \cite{salenbien2012photoacoustic}.
In our experiments, the temperature scanning measurements are performed from 320 K to 200 K with a step of 1 K, under the excitation of three different gratings with $d$ of 10, 14, and 20 \textmu m.\\

\begin{figure*}
\begin{center}
\includegraphics[width=1\textwidth]{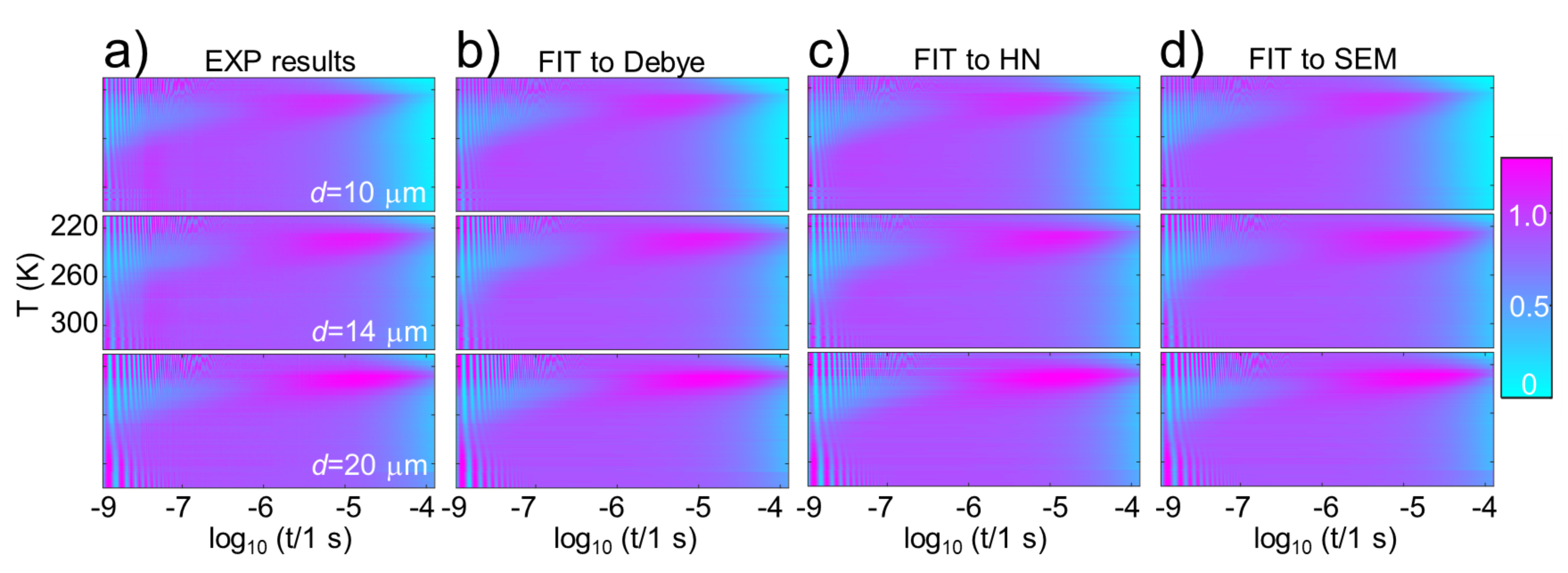} 
\caption{a) Experimental ISS signal (colour scale, arbitrary units) of supercooled glycerol over a broad temperature (vertical axis) and time window (horizontal axis).
The grating size is 10 \textmu m (top), 14 \textmu m (central), and 20 \textmu m (bottom).
b) to d) best fit based on Debye, HN models, and the SEM, respectively. A full presentation of the best fit of all the waveforms is shown in the online Movies.}
\label{results_ISS}
\end{center}
\end{figure*}
Fig. \ref{results_ISS}(a) presents the recorded ISS waveform datasets. As DC-temperature decreases, the acoustic ripples at short times shift the oscillation frequencies from low to high, 60-350 MHz covered by the three gratings, with the attenuation reaching a maximum around 280 K.
This observation reflects the undercooling of the sample, the latter undergoing a transition from liquid-like to glassy-like, and solid-like due to reduced molecular mobility \cite{liu1998jamming}.
The overshoot-like response is noteworthy, spanning from the start of the signal (bluish region), where it overlaps with the acoustic oscillations and fast part of thermal expansion, till the late times (reddish region), when it is quenched by the thermal diffusion dominated part (bluish tail).
This process is the manifestation of the relaxation of heat capacity and thermal expansion coefficient, which are strongly (quasi exponentially) temperature dependent.\\

\begin{figure}[h]
\begin{center}
\includegraphics[width=0.45\textwidth]{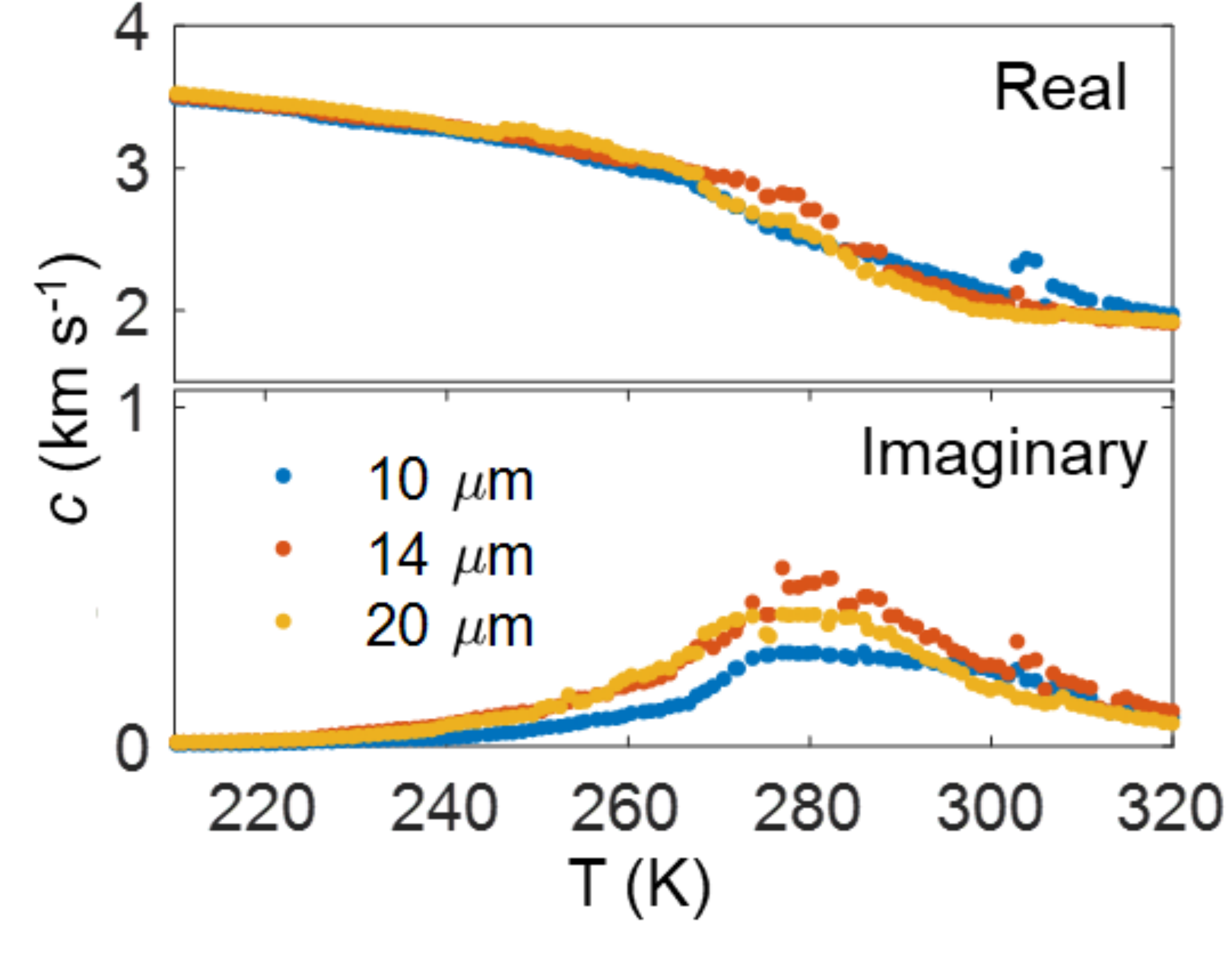} 
\caption{Temperature dependent complex longitudinal velocity ($c$) determined with the three gratings.
The top (bottom) panel corresponds to the real (imaginary) part of $c$.}
\label{V_complex}
\end{center}
\end{figure}

\begin{figure}
\begin{center}
\includegraphics[width=0.45\textwidth]{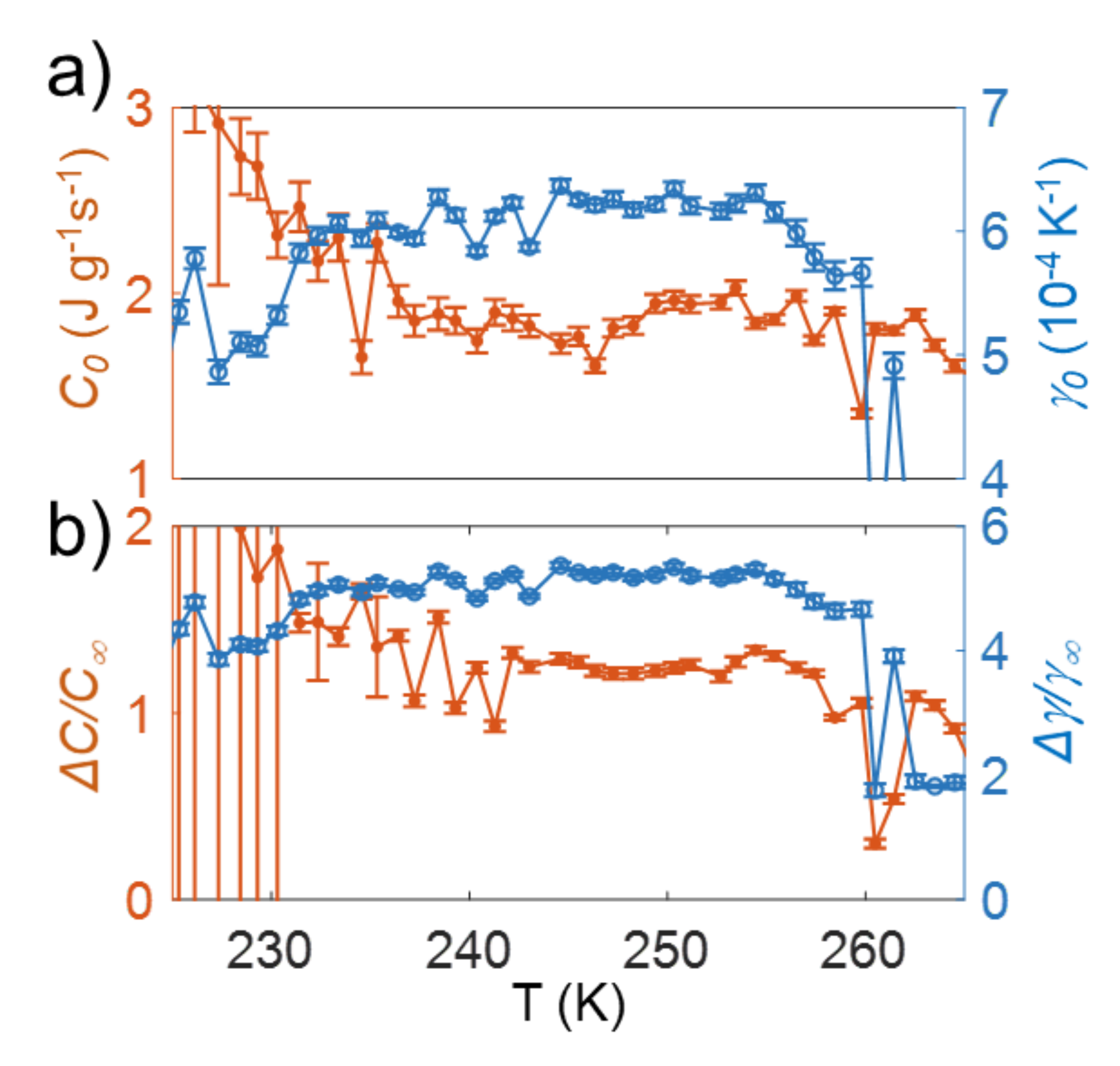} 
\caption{a) low-frequency limit of $C$ (left axis, red) and $\gamma$ (right axis, blue) vs temperatures. b) The relative ratio of each relaxation quantity, $\Delta C$/$C_0$ (left axis, red) and $\Delta\gamma $/$\gamma_0$ (right axis, blue). The data are obtained from the fit in the frame of Debye model and for $d=14$ \textmu m.}
\label{C_gamma_debye_waller}
\end{center}
\end{figure}

A comparative fitting analysis of the acquired ISS datasets is carried out through the two analytical physical models, developed in this work, coupled with Debye and HN relaxation function, and also through the SEM, the latter relying on a single stretched exponential.
A full presentation of the best fits for all signals, obtained with the three models, is summarized in Fig. \ref{results_ISS} (b-d) and also available in the online Movies (1-3) in the Supplemental Material.
Satisfactory fit quality is overall achieved at all temperatures and grating periods by the three models, confirming again the reliability of physical models developed in this work.
ISS signals are information-rich, providing access to the mechanical and thermal relaxation dynamics in a single waveform, which will be discussed in the following.\\
By fitting the experimental traces with our models, for every temperature and light grating we retrieve $c_L$ and $\tau_\eta$.
By inserting these parameters into Eq. \ref{def_c}, we can calculate the complex velocity  of the medium at the acoustic frequency imposed by the grating $\omega_a=2\pi c_L/d$.
Fig. \ref{V_complex} shows the obtained complex sound velocity of the supercooled glycerol at different temperatures determined with the three gratings. The real part of $c$ (top panel) increases upon cooling because of the stiffening of the liquid. The imaginary part of $c$ (bottom panel) reaches a maximum around 280 K, where the structural relaxation timescale overlaps with the acoustic frequency. The results are in good agreement with those reported in Ref. \cite{paolucci2000impulsive}. It is interesting to notice that both the real and imaginary part undergoes a transition around 280 K, which is a reflection of the strong coupling between the acoustic motion and structural changes of the network when $1/f_A$ being of the order of the structural relaxation time.
This feature provides a way to study the mechanical relaxation by performing measurements at numerous grating spacings in a broad range \cite{hecksher2017toward}, namely a mechanical spectroscopic analysis like the traditional rheological spectroscopy \cite{jensen2018slow} or ultrasonic spectroscopy \cite{jeong1986ultrasonic,schroyen2020bulk}.\\
In addition to the mechanical relaxation dynamics, the models developed in this work enable the individual and simultaneous determination of the heat capacity $C$ and the thermal expansion coefficient $\gamma$ relaxation.
\begin{figure}
\begin{center}
\includegraphics[width=0.45\textwidth]{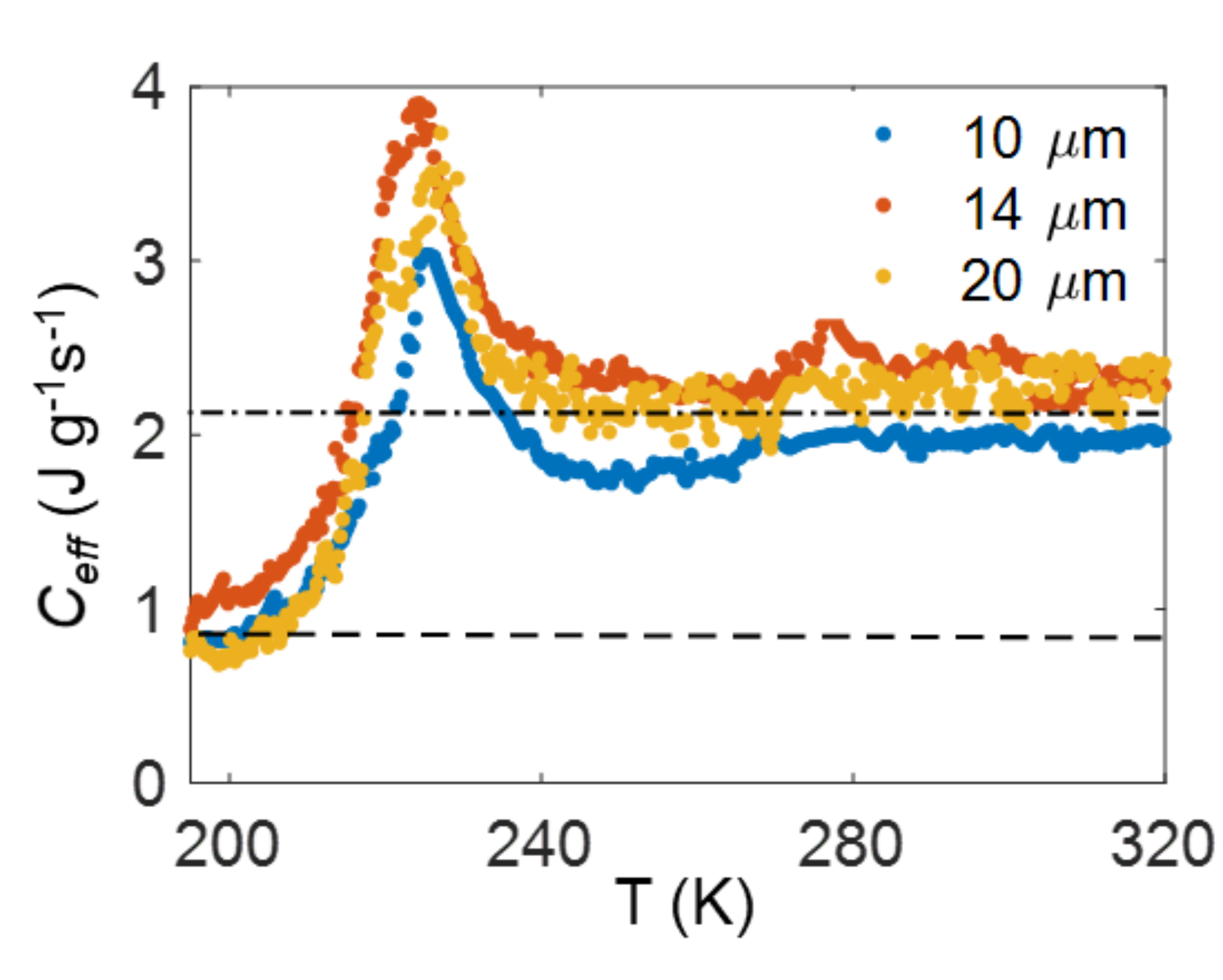} 
\caption{The relaxation of $C$ is also manifested in the thermal diffusion tail of the signal, via its influence on the effective thermal diffusivity, $\alpha_{eff}$=$\kappa/(\rho C_{eff})$. At the low and high temperature limits, the value of $C_{eff}$ extracted from the thermal diffusion tail corresponds well to the respective asymptotic values $C_0$ and $C_{\infty}$, indicated by the dashed lines. }
\label{C_effective}
\end{center}
\end{figure}
In the following, we focus on the case of $d=14$ \textmu m, analyzed in the frame of Debye model, the other cases yielding the same conclusions.
Fig. \ref{C_gamma_debye_waller} (a) shows the obtained low-frequency limit response of $C$ (left axis, red) and $\gamma$ (right axis, blue), in the frame of Debye model.
Panel (b) displays the fitted ratio of $\Delta C/C_{\infty}$ (left axis, red) and $\Delta \gamma/\gamma_{\infty}$ (right axis, blue) at different temperatures.
Within the margin of uncertainty (error bars in Fig. \ref{C_gamma_debye_waller}), determined by the most square error analysis \cite{salenbien2011laser, ThermalLens}, no temperature dependence is observed for all the parameters.
Large fitting uncertainty was found when $T<230$ K and $T>260$ K.
This is because the slow parts of the responses, which are determined by the relaxation strengths and the relaxation frequencies, occur later than 100 \textmu s (when $T<230$ K), and thus after the thermal diffusion driven decay of the signal, or before 1 ns (when $T> 260$ K), the experimentally accessible time window, respectively. We thus used the values between 230-260 K to calculate the average as a representation. Access to lower temperatures can be enabled by using larger grating spacing.
In an accompanying  work, we have demonstrated the use of thermal lens technique \cite{ThermalLens}, with a focused Gaussian beam of about 30 \textmu m, to study the relaxation down to 200 K.\\
The average $C_0$ and $\gamma_0$ from ISS technique are $1980\pm160$ J Kg$^{-1}$ K$^{-1}$ and $(5.5\pm0.7)\times 10^{-4}$ K$^{-1}$, respectively.
The average ratios $\Delta C/C_\infty$ and $\Delta \gamma/\gamma_\infty$ are $1.2\pm0.2$ and $4.9\pm0.7$, for $C$ and $\gamma$, respectively. Using the latter four fitting parameters, one can further calculate the high-frequency limit response, $910\pm150$ J Kg$^{-1}$ K$^{-1}$ for $C_{\infty}$ and $(1.0\pm0.2)\times 10^{-4}$ K$^{-1}$ for $\gamma_{\infty}$, and the relaxation strength ($R_S$), defined as $\Delta C/C_0$ and $\Delta \gamma/\gamma_0$, $0.5\pm0.1$ and $0.81\pm0.04$, respectively.
The obtained results comply well with the data reported in literature, as summarized in Table \ref{table_rel_quan_exp}.\\
By fitting with SEM model, the Debye-Waller factor \cite{paolucci2000impulsive}, $B/(A+B)$ in Eq. \ref{Str_exp_expr}, is used to describe the relaxation strength and we found a value of about $0.65\pm0.05$, which is in good agreement with the one reported in Ref. \cite{paolucci2000impulsive}, the latter reading being $0.66$.
\begin{table}
\begin{center}
\begin{tabular}{|c|c|c|c|c|}\hline
&	Fit	&		3-omega	&	PPE	&	DSC\\ \hline
$C_0$ (J Kg$^{-1}$ K$^{-1}$)	&	$1980\pm160$	&	2071	&	2100	&	2000\\ \hline
$C_\infty$ (J Kg$^{-1}$ K$^{-1}$)	&	$910\pm150$	&	1070	&	1180	&	1000\\ \hline
$R_{S}$ 	&	$0.5\pm0.1$	&	0.48	&	0.44	&	0.5\\ \hline
Ref. 	&	Current work	&	\cite{birge1985specific,birge1986specific}	&	\cite{bentefour2003broadband,bentefour2004thermal}	&	\cite{wang2002direct}\\ \hline
\multicolumn{5}{c}{}\\ \cline{1-3}
	& Fit	&		Dilatometer	& \multicolumn{2}{c}{}\\ \cline{1-3}
$\gamma_0$ ($10^{-4}K^{-1}$)	&	$5.5\pm0.7$	&	1	& \multicolumn{2}{c}{}\\ \cline{1-3}
$\gamma_\infty$ ($10^{-4}$K$^{-1}$) &	$1.0\pm0.2$	&	5	& \multicolumn{2}{c}{}\\ \cline{1-3}
$R_{S}$ 	&	$0.81\pm0.04$	&	0.8	& \multicolumn{2}{c}{}\\ \cline{1-3}
Ref.	&	Current work	&	\cite{blazhnov2004temperature}	& \multicolumn{2}{c}{}\\ \cline{1-3}
\end{tabular}
\end{center}
\caption{Low-frequency and high-frequency limit of the average relaxing quantity $C$ and $\gamma$ and comparison with results in literature obtained with 3-omega, differential scanning calorimetry (DSC), and photopyroelectric spectroscopy (PPE).
In 3-omega and PPE, one measures thermal effusivity ($e$), from which $C(\omega)$ may be indirectly obtained via $e^2=C\kappa_T$, with $\kappa_T$ the thermal conductivity. To perform the conversion, we used $\kappa_T=$0.29 W m$^{-1}$ K$^{-1}$.
}
\label{table_rel_quan_exp}
\end{table}
The value lies in between the relaxation strength of $C$ and $\gamma$, which is expected in the sense that the two relaxing quantities are implicitly incorporated together into a single stretched exponential function.\\
Interestingly, the asymptotic values of the heat capacity can also be extracted, independently of the used models, from the temperature dependence of the thermal diffusion tail of the signal, as depicted in Fig. \ref{ISS_strVsHN}. Provided the relaxation time of the heat capacity and thermal expansion occur before or after the time window of the thermal diffusion tail, the signal tail evolves simply proportional with  $\exp (-q^2\alpha_{eff} t)$ with $\alpha_{eff}$ an effective thermal diffusivity value, connected to the specific heat via $\alpha_{eff}$=$\kappa_T/(\rho C_{eff})$. $\kappa_T$ and $\rho$ denote thermal conductivity and mass density respectively.
In light of their weak temperature dependence \cite{blazhnov2004temperature,minakov2001simultaneous}, in this work the latter two parameters have been assumed as constant, as 0.29 W m$^{-1}$ K$^{-1}$ and 1260 Kg m$^{-3}$, respectively.\\
In Fig. \ref{C_effective} we report $C_{eff}$ as a function of temperature for the three gratings.
The asymptotic values of $C$ for low and high temperatures were found to be $960\pm20$ and $2190\pm30$ J Kg$^{-1}$ K$^{-1}$, as indicated by the two dashed lines, corresponding to a relaxation strength of 0.56, consistent with the value obtained by model fitting, 0.53.
\begin{figure*}
\begin{center}
\includegraphics[width=0.8\textwidth]{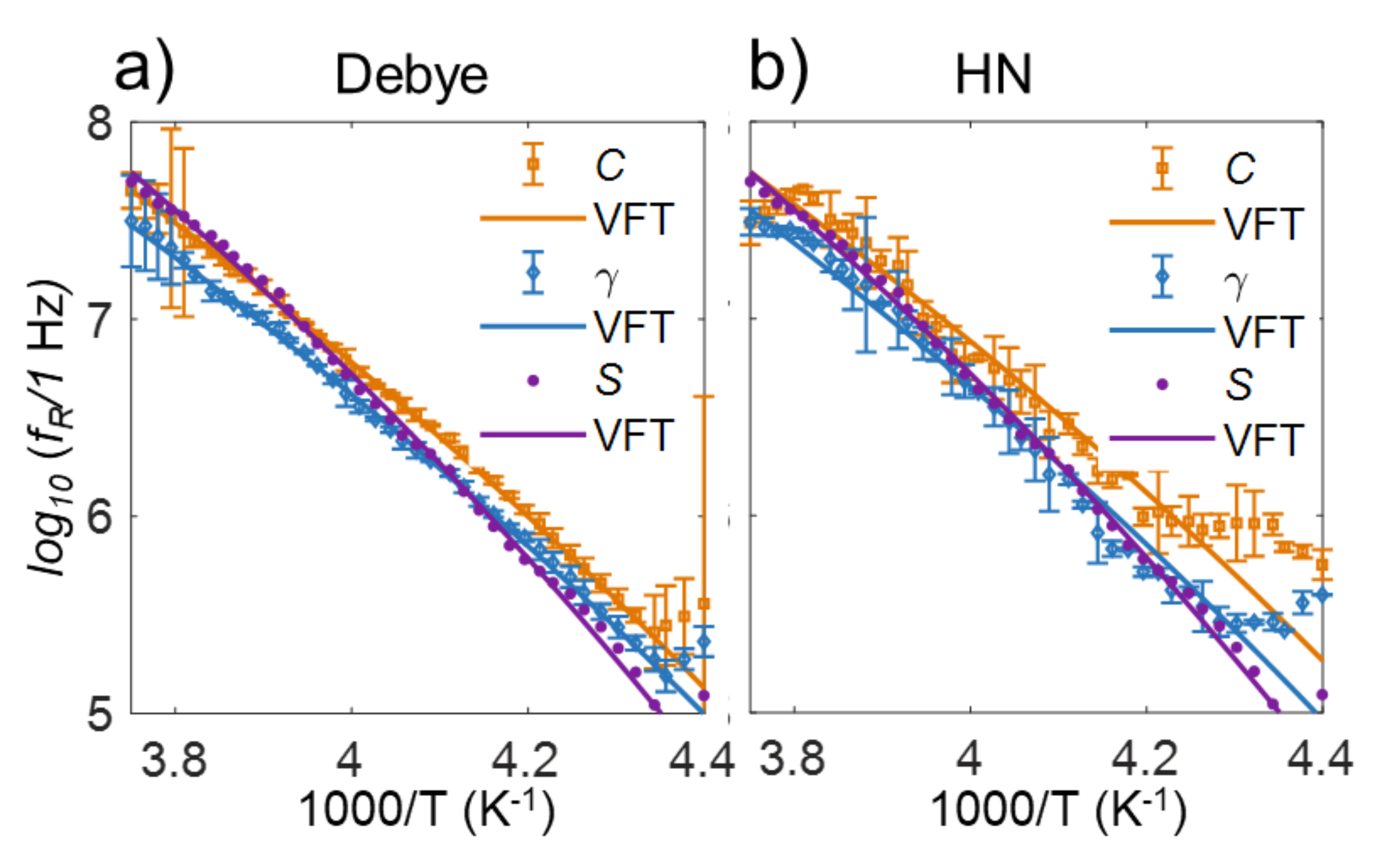} 
\caption{Comparison of the temperature dependent relaxation frequency $f_R$ extracted through Debye model (yellow squares), HN model (blue circles) and SEM (purple circles) and their fit with VFT (solid lines).}
\label{f_relax} 
\end{center}
\end{figure*}
The empirical model assumes that the relaxation for $C$ and $\gamma$ occurring on the same time scale and connects their contribution to the ISS signal into a single stretched exponential function.\\
In order to experimentally verify whether the two response functions are indeed characterized by the same time scale and to what extent they can be disentangled, in Fig. \ref{f_relax} we compare the characteristic relaxation frequency of the heat capacity and of the thermal expansion coefficient (defined as $2\pi/\omega_C$ and $2\pi/\omega_\gamma$, respectively), both in the frame of Debye and Hn models, with the relaxation frequency $\Gamma_R$ of the SEM.
In the case of Debye model (Fig. \ref{f_relax} a), the heat capacity relaxation frequency (yellow squares) is systematically higher (about a factor of $1.5\pm 0.1$) than the one of the thermal expansion coefficient  (blue diamonds). 
This implies that after photothermally supplying energy, first heat is transferred from vibrational energy levels to configurational energy changes and, somewhat later,
\begin{table}
\begin{center}
\begin{tabular}{|c|c|c|}\hline
\multicolumn{3}{|c|}{Heat capacity}\\ \hline
&Debye&HN\\ \hline
$\log_{10}$($f_0/1$ Hz)&14.5 &14.5\\ \hline
$B$ (K)&2140 &2100\\ \hline
$T_0$ (K)&127 &124\\ \hline
$m$ & 50.9 &50.9\\ \hline
\multicolumn{3}{c}{}\\ \hline
\multicolumn{3}{|c|}{Thermal expansion coefficient}\\ \hline
&Debye&HN\\ \hline
$\log_{10}$($f_0/1$ Hz)&13.9 &13.9\\ \hline
$B$ (K)&2011 &2195\\ \hline
$T_0$ (K)&130 &125\\ \hline
$m$ & 54.1 &49.9\\ \hline
\multicolumn{3}{c}{}\\ \cline{1-2}
\multicolumn{2}{|c|}{SEM}&\multicolumn{1}{l}{}\\ \cline{1-2}
$\log_{10}$($f_0/1$ Hz)&14.8&\multicolumn{1}{l}{}\\ \cline{1-2}
$B$ (K)&2138&\multicolumn{1}{l}{}\\ \cline{1-2}
$T_0$ (K)&135&\multicolumn{1}{l}{}\\ \cline{1-2}
\end{tabular}
\end{center}
\caption{Summary of the parameters extracted by fitting with the VFT expression.
}
\label{table_VFT_parameters}
\end{table}
the configurational energy changes result in an increase of volume, in agreement with the two-temperature model developed in Section \ref{Sec:two_temp_model}.\\
Similar conclusions can be drawn from the results obtained by the HN model as shown in Fig. \ref{f_relax} (b). However, the results from HN model fitting are more dispersed due to the co-variance with the additional two fitting variables, namely a and b in Eq. \ref{C_omega_HN}.\\
The structural relaxation frequency $\Gamma_R$ (purple circles), obtained by fitting the experimental data with the SEM model,  characterizes the (combined thermal and thermal expansion) structural relaxation and turn out to lie somewhat in between the other two relaxation frequencies.\\
The obtained temperature dependence of the relaxation frequencies were fitted to the Vogel-Fulcher-Tamman (VFT) equation (solid lines in Fig. \ref{f_relax}), defined by $f_{relax}=f_0\exp\left[-B/(T-T_0)\right]$, with $f_0$ the relaxation frequency in the high temperature limit, $T_0$ the Vogel-Fulcher temperature, around 130 K for glycerol.
In Table \ref{table_VFT_parameters} we report the fitted VFT parameters based on Debye, HN and SEM. The results for the latter model are in line with values reported in Ref. \cite{paolucci2000impulsive}.\\
From the ratio $D=B/T_0$ we have determined the so-called fragility $m$, via $m= 16+590/D$, which can be considered as a measure for the deviation from Arrhenius behavior (and thus as a measure for the degree of temperature dependence of the potential energy landscape morphology).
The fragility values obtained with our model are summarized in Table \ref{table_VFT_parameters} and are close to 53, the latter being the fragility for glycerol reported in Ref. \cite{bohmer1993nonexponential}.\\

\section{conclusion}
In this manuscript, a model to describe ISS signals generated in relaxing materials has been introduced, which is based on the solution of the thermal diffusion equation and the continuum mechanics equation, in combination with a frequency dependent heat capacity and thermal expansion coefficient. As functional forms for the frequency dependencies, Debye and Havriliak-Negami expressions were assumed. The assumption of a Debye frequency dependence of the heat capacity was shown to be compatible with a two-temperature model, in which the experimentally measured temperature refers to the energy distribution of the kinetic degrees of freedom, and a network temperature describes the state of the amorphous network, which is assumed to be in thermal contact with the kinetic energy reservoir.\\
The obtained physical models for describing ISS response, were shown to fit well ISS signals that had been simulated, for different temperature-wavenumber combinations in glycerol, by a semi-empirical model \cite{yang1995impulsive1} that has been historically used to describe ISS signals in relaxing materials.\\ Furthermore, we have carried out an experimental ISS investigation of glycerol under supercooling and also a comparative model fitting analysis based the physical models developed in this work and the existing empirical model. Satisfied fitting quality has been  achieved for all ISS waveforms, confirming the models developed in this work and allowing us to study the relaxation of $C$ and $\gamma$, up to several tens of MHz, largely extending the upper limit of spectroscopy of thermal susceptibility, by nearly 3 and 7 decades for $C(\omega)$ and $\gamma(\omega)$, respectively. The best fit results also suggest that the relaxation of heat capacity and thermal expansivity occur on a slight different time scale, relaxation of $C$ is about 1.5 times faster than that of $\gamma$, which is line with the observation by a thermal lens spectroscopy investigation, reported in an accompanying article \cite{ThermalLens}.

\appendix

\section{Derivation of the expression for $\tilde{T}_N$}
\label{app:Derivation of the}
\noindent In this appendix we derive a simple analytical expression linking the network and the KER's temperatures in frequency domain.
The starting point is the equation linking $\tilde{T}_N$ to $\tilde{T}$ and its derivatives, obtained by Fourier transforming \ref{first_TTM_rewritten}:
\begin{equation}
\tilde{T}_N(x,\omega)=\tilde{T}+\frac{\rho C}{G}i\omega \tilde{T}-\frac{\kappa_T}{G}\frac{\partial^2 \tilde{T}}{\partial x^2}-{1\over G}\tilde{Q}(x,\omega).
\label{tildeTN_transformed}
\end{equation}
In order to simplify the latter equation, we need to write 
$\partial^2\tilde{T}/\partial x^2$ in terms of $\tilde{T}$.
To this purpose, we write general solution of Eq. \ref{decoupled_for_T_freq_4} as:
\begin{equation}
\tilde{T}(x,\omega)=E_1(\omega)\exp\left(+\hat{\zeta} x\right)+E_2(\omega)\exp\left(-\hat{\zeta} x\right)+y_p(x,\omega),
\label{T_omega_system}
\end{equation}
where $\hat{\zeta}$ is defined as:
\begin{equation}
\hat{\zeta}^2= \frac{i\omega \rho}{\kappa_T}\left[C+\frac{{\rho_N \over \rho} C_N}{1+i\frac{\rho_N C_N}{G}\omega}\right].
\label{def_of_hat_tau}
\end{equation}
$y_p(x,\omega)$ is a particular solution of Eq. \ref{decoupled_for_T_freq_4}.
The function, 
$$y_p(x,\omega)={1\over 2\hat{\zeta} \kappa_T}\left[-\tilde{H}_1(x,\omega)\exp\left(+\hat{\zeta} x\right)\right.$$
\begin{equation}
\left.+\tilde{H}_2(x,\omega)\exp\left(-\hat{\zeta} x\right)\right],
\label{y_part_for_T_with_H}
\end{equation}
with $\tilde{H}_1$ and $\tilde{H}_2$ satisfying the following conditions:
\begin{equation}
\frac{\partial \tilde{H}_1}{\partial x}=\tilde{Q}(x,\omega)\exp\left(-\hat{\zeta} x\right),
\end{equation}
\begin{equation}
\frac{\partial\tilde{H}_2}{\partial x}=\tilde{Q}(x,\omega)\exp\left(+\hat{\zeta} x\right).
\end{equation}
is a particular solution of Eq. \ref{decoupled_for_T_freq_4}. To prove this assertion, we first calculate the first spatial derivatives of $y_p(x,\omega)$:
\begin{equation}
\frac{\partial y_p}{\partial x}={1\over 2 \kappa_T}\left[-\tilde{H}_1(x,\omega)\exp\left(+\hat{\zeta} x\right)-\tilde{H}_2(x,\omega)\exp\left(-\hat{\zeta} x\right)\right].
\end{equation}
Consequently, the second spatial derivative reads
\begin{equation}
\frac{\partial y_p^2}{\partial x^2}=\hat{\zeta}^2 y_p(x,\omega) -{1\over \kappa_T}\tilde{Q}(x,\omega).
\label{yp_T_particular_solution}
\end{equation}
The latter expression proves that $y_p(x,\omega)$ is the particular solution of Eq. \ref{decoupled_for_T_freq_4} we were looking for.\\
Accordingly, the second spatial derivative of the KER's temperature reads:
$$\frac{\partial^2 \tilde{T}}{\partial x^2}=\hat{\zeta}^2\left[E_1(\omega)\exp\left(+\hat{\zeta} x\right)+E_2(\omega)\exp\left(-\hat{\zeta} x\right)\right]+\frac{\partial^2 y_p}{\partial x^2}=$$
$$=\hat{\zeta}^2\left[E_1(\omega)\exp\left(+\hat{\zeta} x\right)+E_2(\omega)\exp\left(-\hat{\zeta} x\right)\right]+$$
$$+\hat{\zeta}^2 y_p(x,\omega) -{1\over \kappa_T}\tilde{Q}(x,\omega)=$$
\begin{equation}
=\hat{\zeta}^2\tilde{T}(x,\omega) -{1\over \kappa_T}\tilde{Q}(x,\omega),
\label{second_derivative_of_T_omega}
\end{equation}
where the second equality followed from Eq. \ref{yp_T_particular_solution}.\\
We can now also derive the network's temperature.
Finally, by substituting Eq. \ref{second_derivative_of_T_omega} into \ref{tildeTN_transformed} and recalling the definition of $\hat{\zeta}$ we obtain:
$$\tilde{T}_N(x,\omega)=\left(1+\frac{\rho C}{G}i\omega-\frac{\kappa_T}{G}\hat{\zeta}^2\right)\tilde{T}=$$
\begin{equation}
=\left(\frac{1}{1+i\frac{\rho_N C_N}{G}\omega}\right)\tilde{T}(x,\omega).
\label{T_N_vs_T}
\end{equation}

\section{Application of residue theorem to evaluate inverse Fourier's transform}
\label{appen:res_th}
We consider the function:
$$F(\omega)=A\frac{\prod_{j=1}^{N}(\omega-z_j)}{\prod_{k=1}^{M}(\omega-p_k)^{m_k}},$$
where $A$ is a constant, $m_k$ are real positive numbers, $z_j$ are complex numbers, $p_k$ are complex numbers with strictly positive imaginary part. We assume that the degree of the denominator is greater then that of the numerator.
Hence, $F(\omega)$ has $M$ poles falling in the positive part of the complex plan.\\
The Fourier's inverse transform of $F(\omega)$ reads:
$$f(t)=\int_{-\infty}^{+\infty} F(\omega)\exp(i\omega t)d\omega.$$
For $t<0$,
$$f(t)=\int_{-\infty}^{+\infty}F(\omega)\exp(i\omega t)d\omega=$$
$$=\int_{-\infty}^{+\infty}F(\omega)\exp(i\omega t)\omega+\int_{\gamma_1}F(\omega)\exp(i\omega t)d\omega=$$
$$=\oint_{\Gamma_1}F(\omega)\exp(i\omega t)d\omega=0.$$

$\gamma_1$ is a semicircle parametrized with the coordinates $\omega=R(\cos\theta-i\sin \theta)$, with $R\rightarrow+\infty$ and $\theta$ running from 0 to $\pi$. In other words, $\gamma_1$ is a semicircle in the clockwise direction, located in the lower part of the complex plane and with very big radius.
For the Jordan's lemma, the integral of $F(\omega)\exp(i\omega t)$ over $\gamma_1$ is 0.
$\Gamma_1=\gamma_1\cup[-\infty,+\infty]$ is the closed path over which the residue theorem is evaluated. The direction of $\Gamma_1$ is clockwise. $\Gamma_1$ does not surround any poles, hence the integral over $\Gamma_1$ gives 0.
This yields $f(t)=0$ for $t<0$.\\
In analogy, for $t>0$,
$$f(t)=\int_{-\infty}^{+\infty}F(\omega)\exp(i\omega t)d\omega=$$
$$=\int_{-\infty}^{+\infty}F(\omega)\exp(i\omega t)d\omega+\int_{\gamma_2}F(\omega)\exp(i\omega t)d\omega=$$
$$=\oint_{\Gamma_2}F(\omega)\exp(i\omega t)d\omega=2\pi i\sum_{k=1}^{M} Res(p_k).$$
$Res(p_k)$ is the residue corresponding to the $k$-th pole (of order $m_k$), calculated as:
$$Res(p_k)=$$
$$=\frac{1}{\left(m_k-1\right)!}\lim_{\omega\rightarrow p_k}\frac{d^{m_k-1}}{d\omega^{m_k-1}}\left\{F(\omega)\exp(i\omega t)(\omega-p_k)^{m_k}\right\}.$$
$\gamma_2$ is a semicircle parametrized with the coordinates $\omega=R(\cos\theta+i\sin \theta)$, with $R\rightarrow+\infty$ and $\theta$ running from 0 to $\pi$.
Jordan's lemma indicates that the integral of $F(\omega)\exp(i\omega t)$ over $\gamma_2$ is 0.
$\Gamma_2=\gamma_2\cup[-\infty,+\infty]$ is the closed path over which the residue theorem is evaluated. The direction of $\Gamma_2$ is counterclockwise.\\
Finally we have to consider the case $t=0$.
Before considering the integral, we use the fact that the degree of the denominator of $F$ exceeds that of the numerator. Hence, we can rewrite $F$ with the partial fraction decomposition:
$$F(\omega)=\sum_{k=1}^M\sum_{j=1}^{m_k}\frac{a_{kj}}{(\omega-p_k)^j},$$
where $a_{kj}$ are complex numbers.\\
The integral to be computed yields:
$$f(t=0)=\int_{-\infty}^{+\infty} F(\omega)d\omega=\sum_{k=1}^M\sum_{j=1}^{m_k}a_{kj}\int_{-\infty}^{+\infty} \frac{1}{(\omega-p_k)^j}d\omega.$$
When $j=1$, we have that:
$$\int_{-\infty}^{+\infty} \frac{1}{(\omega-p_k)}d\omega=\left[\ln|\omega-p_k|\right]_{-\infty}^{+\infty}=0.$$
On the other hand, when $j>1$ we have:
$$\int_{-\infty}^{+\infty} \frac{1}{(\omega-p_k)^j}d\omega=\left[\frac{1}{(1-j)(\omega-p_k)^{j-1}}\right]_{-\infty}^{+\infty}=0.$$
For this reason, $f(t=0)=0$.\\
Putting everything together, the general expression of $f(t)$ reads:
\begin{equation}
f(t)=2\pi i\sum_{k=1}^{M} Res(p_k)\theta(t).
\label{Fourier_transform_inverse_of_F}
\end{equation}

\section*{Acknowledgements}
CG and MG are grateful to the KU Leuven Research Council for financial support (C14/16/063 OPTIPROBE).
MG acknowledges financial support from the CNR Joint Laboratories program 2019-2021, project SAC.AD002.026 (OMEN).
LL acknowledges the financial support from FWO (Research Foundation-Flanders) postdoctoral research fellowship (12V4419N).
PZ acknowledges the support of Chinese Scholarship Council (CSC).
F. B. acknowledges financial support from Universit\'{e} de Lyon in the frame of the IDEXLYON Project (ANR-16-IDEX-0005) and from Universit\'{e} Claude Bernard Lyon 1 through the BQR Accueil EC 2019 grant.

\bibliography{bibliography}

\begin{thebibliography}{48}
\expandafter\ifx\csname natexlab\endcsname\relax\def\natexlab#1{#1}\fi
\expandafter\ifx\csname bibnamefont\endcsname\relax
  \def\bibnamefont#1{#1}\fi
\expandafter\ifx\csname bibfnamefont\endcsname\relax
  \def\bibfnamefont#1{#1}\fi
\expandafter\ifx\csname citenamefont\endcsname\relax
  \def\citenamefont#1{#1}\fi
\expandafter\ifx\csname url\endcsname\relax
  \def\url#1{\texttt{#1}}\fi
\expandafter\ifx\csname urlprefix\endcsname\relax\def\urlprefix{URL }\fi
\providecommand{\bibinfo}[2]{#2}
\providecommand{\eprint}[2][]{\url{#2}}

\bibitem[{\citenamefont{Yang and
  Nelson}(1995{\natexlab{a}})}]{yang1995impulsive1}
\bibinfo{author}{\bibfnamefont{Y.}~\bibnamefont{Yang}} \bibnamefont{and}
  \bibinfo{author}{\bibfnamefont{K.~A.} \bibnamefont{Nelson}},
  \bibinfo{journal}{The Journal of chemical physics}
  \textbf{\bibinfo{volume}{103}}, \bibinfo{pages}{7722}
  (\bibinfo{year}{1995}{\natexlab{a}}).

\bibitem[{\citenamefont{Bapst et~al.}(2020)\citenamefont{Bapst, Keck,
  Grabska-Barwi{\'n}ska, Donner, Cubuk, Schoenholz, Obika, Nelson, Back,
  Hassabis et~al.}}]{bapst2020unveiling}
\bibinfo{author}{\bibfnamefont{V.}~\bibnamefont{Bapst}},
  \bibinfo{author}{\bibfnamefont{T.}~\bibnamefont{Keck}},
  \bibinfo{author}{\bibfnamefont{A.}~\bibnamefont{Grabska-Barwi{\'n}ska}},
  \bibinfo{author}{\bibfnamefont{C.}~\bibnamefont{Donner}},
  \bibinfo{author}{\bibfnamefont{E.~D.} \bibnamefont{Cubuk}},
  \bibinfo{author}{\bibfnamefont{S.~S.} \bibnamefont{Schoenholz}},
  \bibinfo{author}{\bibfnamefont{A.}~\bibnamefont{Obika}},
  \bibinfo{author}{\bibfnamefont{A.~W.} \bibnamefont{Nelson}},
  \bibinfo{author}{\bibfnamefont{T.}~\bibnamefont{Back}},
  \bibinfo{author}{\bibfnamefont{D.}~\bibnamefont{Hassabis}},
  \bibnamefont{et~al.}, \bibinfo{journal}{Nature Physics}
  \textbf{\bibinfo{volume}{16}}, \bibinfo{pages}{448} (\bibinfo{year}{2020}).

\bibitem[{\citenamefont{Jensen et~al.}(2018)\citenamefont{Jensen, Gainaru,
  Alba-Simionesco, Hecksher, and Niss}}]{jensen2018slow}
\bibinfo{author}{\bibfnamefont{M.~H.} \bibnamefont{Jensen}},
  \bibinfo{author}{\bibfnamefont{C.}~\bibnamefont{Gainaru}},
  \bibinfo{author}{\bibfnamefont{C.}~\bibnamefont{Alba-Simionesco}},
  \bibinfo{author}{\bibfnamefont{T.}~\bibnamefont{Hecksher}}, \bibnamefont{and}
  \bibinfo{author}{\bibfnamefont{K.}~\bibnamefont{Niss}},
  \bibinfo{journal}{Physical Chemistry Chemical Physics}
  \textbf{\bibinfo{volume}{20}}, \bibinfo{pages}{1716} (\bibinfo{year}{2018}).

\bibitem[{\citenamefont{Hecksher et~al.}(2017)\citenamefont{Hecksher,
  Torchinsky, Klieber, Johnson, Dyre, and Nelson}}]{hecksher2017toward}
\bibinfo{author}{\bibfnamefont{T.}~\bibnamefont{Hecksher}},
  \bibinfo{author}{\bibfnamefont{D.~H.} \bibnamefont{Torchinsky}},
  \bibinfo{author}{\bibfnamefont{C.}~\bibnamefont{Klieber}},
  \bibinfo{author}{\bibfnamefont{J.~A.} \bibnamefont{Johnson}},
  \bibinfo{author}{\bibfnamefont{J.~C.} \bibnamefont{Dyre}}, \bibnamefont{and}
  \bibinfo{author}{\bibfnamefont{K.~A.} \bibnamefont{Nelson}},
  \bibinfo{journal}{Proceedings of the National Academy of Sciences}
  \textbf{\bibinfo{volume}{114}}, \bibinfo{pages}{8710} (\bibinfo{year}{2017}).

\bibitem[{\citenamefont{Klieber et~al.}(2013)\citenamefont{Klieber, Hecksher,
  Pezeril, Torchinsky, Dyre, and Nelson}}]{klieber2013mechanical}
\bibinfo{author}{\bibfnamefont{C.}~\bibnamefont{Klieber}},
  \bibinfo{author}{\bibfnamefont{T.}~\bibnamefont{Hecksher}},
  \bibinfo{author}{\bibfnamefont{T.}~\bibnamefont{Pezeril}},
  \bibinfo{author}{\bibfnamefont{D.~H.} \bibnamefont{Torchinsky}},
  \bibinfo{author}{\bibfnamefont{J.~C.} \bibnamefont{Dyre}}, \bibnamefont{and}
  \bibinfo{author}{\bibfnamefont{K.~A.} \bibnamefont{Nelson}},
  \bibinfo{journal}{The Journal of Chemical Physics}
  \textbf{\bibinfo{volume}{138}}, \bibinfo{pages}{12A544}
  (\bibinfo{year}{2013}).

\bibitem[{\citenamefont{Blazhnov et~al.}(2004)\citenamefont{Blazhnov, Malomuzh,
  and Lishchuk}}]{blazhnov2004temperature}
\bibinfo{author}{\bibfnamefont{I.~V.} \bibnamefont{Blazhnov}},
  \bibinfo{author}{\bibfnamefont{N.~P.} \bibnamefont{Malomuzh}},
  \bibnamefont{and} \bibinfo{author}{\bibfnamefont{S.~V.}
  \bibnamefont{Lishchuk}}, \bibinfo{journal}{The Journal of chemical physics}
  \textbf{\bibinfo{volume}{121}}, \bibinfo{pages}{6435} (\bibinfo{year}{2004}).

\bibitem[{\citenamefont{Niss and Hecksher}(2018)}]{niss2018perspective}
\bibinfo{author}{\bibfnamefont{K.}~\bibnamefont{Niss}} \bibnamefont{and}
  \bibinfo{author}{\bibfnamefont{T.}~\bibnamefont{Hecksher}},
  \bibinfo{journal}{The Journal of chemical physics}
  \textbf{\bibinfo{volume}{149}}, \bibinfo{pages}{230901}
  (\bibinfo{year}{2018}).

\bibitem[{\citenamefont{Klieber et~al.}(2015)\citenamefont{Klieber, Gusev,
  Pezeril, and Nelson}}]{klieber2015nonlinear}
\bibinfo{author}{\bibfnamefont{C.}~\bibnamefont{Klieber}},
  \bibinfo{author}{\bibfnamefont{V.~E.} \bibnamefont{Gusev}},
  \bibinfo{author}{\bibfnamefont{T.}~\bibnamefont{Pezeril}}, \bibnamefont{and}
  \bibinfo{author}{\bibfnamefont{K.~A.} \bibnamefont{Nelson}},
  \bibinfo{journal}{Physical Review Letters} \textbf{\bibinfo{volume}{114}},
  \bibinfo{pages}{065701} (\bibinfo{year}{2015}).

\bibitem[{\citenamefont{Gundermann et~al.}(2011)\citenamefont{Gundermann,
  Pedersen, Hecksher, Bailey, Jakobsen, Christensen, Olsen, Schr{\o}der,
  Fragiadakis, Casalini et~al.}}]{gundermann2011predicting}
\bibinfo{author}{\bibfnamefont{D.}~\bibnamefont{Gundermann}},
  \bibinfo{author}{\bibfnamefont{U.~R.} \bibnamefont{Pedersen}},
  \bibinfo{author}{\bibfnamefont{T.}~\bibnamefont{Hecksher}},
  \bibinfo{author}{\bibfnamefont{N.~P.} \bibnamefont{Bailey}},
  \bibinfo{author}{\bibfnamefont{B.}~\bibnamefont{Jakobsen}},
  \bibinfo{author}{\bibfnamefont{T.}~\bibnamefont{Christensen}},
  \bibinfo{author}{\bibfnamefont{N.~B.} \bibnamefont{Olsen}},
  \bibinfo{author}{\bibfnamefont{T.~B.} \bibnamefont{Schr{\o}der}},
  \bibinfo{author}{\bibfnamefont{D.}~\bibnamefont{Fragiadakis}},
  \bibinfo{author}{\bibfnamefont{R.}~\bibnamefont{Casalini}},
  \bibnamefont{et~al.}, \bibinfo{journal}{Nature Physics}
  \textbf{\bibinfo{volume}{7}}, \bibinfo{pages}{816} (\bibinfo{year}{2011}).

\bibitem[{\citenamefont{Glorieux et~al.}(2002)\citenamefont{Glorieux, Nelson,
  Hinze, and Fayer}}]{glorieux2002thermal}
\bibinfo{author}{\bibfnamefont{C.}~\bibnamefont{Glorieux}},
  \bibinfo{author}{\bibfnamefont{K.}~\bibnamefont{Nelson}},
  \bibinfo{author}{\bibfnamefont{G.}~\bibnamefont{Hinze}}, \bibnamefont{and}
  \bibinfo{author}{\bibfnamefont{M.}~\bibnamefont{Fayer}},
  \bibinfo{journal}{The Journal of chemical physics}
  \textbf{\bibinfo{volume}{116}}, \bibinfo{pages}{3384} (\bibinfo{year}{2002}).

\bibitem[{\citenamefont{Silence et~al.}(1992)\citenamefont{Silence, Duggal,
  Dhar, and Nelson}}]{silence1992structural}
\bibinfo{author}{\bibfnamefont{S.~M.} \bibnamefont{Silence}},
  \bibinfo{author}{\bibfnamefont{A.~R.} \bibnamefont{Duggal}},
  \bibinfo{author}{\bibfnamefont{L.}~\bibnamefont{Dhar}}, \bibnamefont{and}
  \bibinfo{author}{\bibfnamefont{K.~A.} \bibnamefont{Nelson}},
  \bibinfo{journal}{The Journal of chemical physics}
  \textbf{\bibinfo{volume}{96}}, \bibinfo{pages}{5448} (\bibinfo{year}{1992}).

\bibitem[{\citenamefont{Yang and Nelson}(1995{\natexlab{b}})}]{yang1995t}
\bibinfo{author}{\bibfnamefont{Y.}~\bibnamefont{Yang}} \bibnamefont{and}
  \bibinfo{author}{\bibfnamefont{K.~A.} \bibnamefont{Nelson}},
  \bibinfo{journal}{Physical review letters} \textbf{\bibinfo{volume}{74}},
  \bibinfo{pages}{4883} (\bibinfo{year}{1995}{\natexlab{b}}).

\bibitem[{\citenamefont{Yang and
  Nelson}(1995{\natexlab{c}})}]{yang1995impulsive2}
\bibinfo{author}{\bibfnamefont{Y.}~\bibnamefont{Yang}} \bibnamefont{and}
  \bibinfo{author}{\bibfnamefont{K.~A.} \bibnamefont{Nelson}},
  \bibinfo{journal}{The Journal of chemical physics}
  \textbf{\bibinfo{volume}{103}}, \bibinfo{pages}{7732}
  (\bibinfo{year}{1995}{\natexlab{c}}).

\bibitem[{\citenamefont{Paolucci and Nelson}(2000)}]{paolucci2000impulsive}
\bibinfo{author}{\bibfnamefont{D.~M.} \bibnamefont{Paolucci}} \bibnamefont{and}
  \bibinfo{author}{\bibfnamefont{K.~A.} \bibnamefont{Nelson}},
  \bibinfo{journal}{The Journal of Chemical Physics}
  \textbf{\bibinfo{volume}{112}}, \bibinfo{pages}{6725} (\bibinfo{year}{2000}).

\bibitem[{\citenamefont{Halalay and
  Nelson}(1992{\natexlab{a}})}]{halalay1992liquid}
\bibinfo{author}{\bibfnamefont{I.}~\bibnamefont{Halalay}} \bibnamefont{and}
  \bibinfo{author}{\bibfnamefont{K.~A.} \bibnamefont{Nelson}},
  \bibinfo{journal}{The Journal of chemical physics}
  \textbf{\bibinfo{volume}{97}}, \bibinfo{pages}{3557}
  (\bibinfo{year}{1992}{\natexlab{a}}).

\bibitem[{\citenamefont{Halalay and
  Nelson}(1992{\natexlab{b}})}]{halalay1992time}
\bibinfo{author}{\bibfnamefont{I.}~\bibnamefont{Halalay}} \bibnamefont{and}
  \bibinfo{author}{\bibfnamefont{K.~A.} \bibnamefont{Nelson}},
  \bibinfo{journal}{Physical review letters} \textbf{\bibinfo{volume}{69}},
  \bibinfo{pages}{636} (\bibinfo{year}{1992}{\natexlab{b}}).

\bibitem[{\citenamefont{Silence et~al.}(1990)\citenamefont{Silence, Goates, and
  Nelson}}]{silence1990impulsive}
\bibinfo{author}{\bibfnamefont{S.~M.} \bibnamefont{Silence}},
  \bibinfo{author}{\bibfnamefont{S.~R.} \bibnamefont{Goates}},
  \bibnamefont{and} \bibinfo{author}{\bibfnamefont{K.~A.}
  \bibnamefont{Nelson}}, \bibinfo{journal}{Chemical physics}
  \textbf{\bibinfo{volume}{149}}, \bibinfo{pages}{233} (\bibinfo{year}{1990}).

\bibitem[{\citenamefont{Birge and Nagel}(1985)}]{birge1985specific}
\bibinfo{author}{\bibfnamefont{N.~O.} \bibnamefont{Birge}} \bibnamefont{and}
  \bibinfo{author}{\bibfnamefont{S.~R.} \bibnamefont{Nagel}},
  \bibinfo{journal}{Physical Review Letters} \textbf{\bibinfo{volume}{54}},
  \bibinfo{pages}{2674} (\bibinfo{year}{1985}).

\bibitem[{\citenamefont{Bentefour et~al.}(2003)\citenamefont{Bentefour,
  Glorieux, Chirtoc, and Thoen}}]{bentefour2003broadband}
\bibinfo{author}{\bibfnamefont{E.~H.} \bibnamefont{Bentefour}},
  \bibinfo{author}{\bibfnamefont{C.}~\bibnamefont{Glorieux}},
  \bibinfo{author}{\bibfnamefont{M.}~\bibnamefont{Chirtoc}}, \bibnamefont{and}
  \bibinfo{author}{\bibfnamefont{J.}~\bibnamefont{Thoen}},
  \bibinfo{journal}{Journal of applied physics} \textbf{\bibinfo{volume}{93}},
  \bibinfo{pages}{9610} (\bibinfo{year}{2003}).

\bibitem[{\citenamefont{Bentefour et~al.}(2004)\citenamefont{Bentefour,
  Glorieux, Chirtoc, and Thoen}}]{bentefour2004thermal}
\bibinfo{author}{\bibfnamefont{E.~H.} \bibnamefont{Bentefour}},
  \bibinfo{author}{\bibfnamefont{C.}~\bibnamefont{Glorieux}},
  \bibinfo{author}{\bibfnamefont{M.}~\bibnamefont{Chirtoc}}, \bibnamefont{and}
  \bibinfo{author}{\bibfnamefont{J.}~\bibnamefont{Thoen}},
  \bibinfo{journal}{The Journal of chemical physics}
  \textbf{\bibinfo{volume}{120}}, \bibinfo{pages}{3726} (\bibinfo{year}{2004}).

\bibitem[{\citenamefont{Niss et~al.}(2012)\citenamefont{Niss, Gundermann,
  Christensen, and Dyre}}]{niss2012dynamic}
\bibinfo{author}{\bibfnamefont{K.}~\bibnamefont{Niss}},
  \bibinfo{author}{\bibfnamefont{D.}~\bibnamefont{Gundermann}},
  \bibinfo{author}{\bibfnamefont{T.}~\bibnamefont{Christensen}},
  \bibnamefont{and} \bibinfo{author}{\bibfnamefont{J.~C.} \bibnamefont{Dyre}},
  \bibinfo{journal}{Physical Review E} \textbf{\bibinfo{volume}{85}},
  \bibinfo{pages}{041501} (\bibinfo{year}{2012}).

\bibitem[{\citenamefont{Liu et~al.}(2021)\citenamefont{Liu, Gandolfi,
  Salenbien, Banfi, and Glorieux}}]{Liu2021}
\bibinfo{author}{\bibfnamefont{L.}~\bibnamefont{Liu}},
  \bibinfo{author}{\bibfnamefont{M.}~\bibnamefont{Gandolfi}},
  \bibinfo{author}{\bibfnamefont{R.}~\bibnamefont{Salenbien}},
  \bibinfo{author}{\bibfnamefont{F.}~\bibnamefont{Banfi}}, \bibnamefont{and}
  \bibinfo{author}{\bibfnamefont{C.}~\bibnamefont{Glorieux}},
  \bibinfo{journal}{submitted to Physical Review Letters}
  (\bibinfo{year}{2021}).

\bibitem[{\citenamefont{Zhang et~al.}(2021)\citenamefont{Zhang, Liu, Gandolfi,
  and Glorieux}}]{ThermalLens}
\bibinfo{author}{\bibfnamefont{P.}~\bibnamefont{Zhang}},
  \bibinfo{author}{\bibfnamefont{L.}~\bibnamefont{Liu}},
  \bibinfo{author}{\bibfnamefont{M.}~\bibnamefont{Gandolfi}}, \bibnamefont{and}
  \bibinfo{author}{\bibfnamefont{C.}~\bibnamefont{Glorieux}},
  \bibinfo{journal}{submitted to Physical Review B}  (\bibinfo{year}{2021}).

\bibitem[{\citenamefont{Gandolfi et~al.}(2019)\citenamefont{Gandolfi, Benetti,
  Glorieux, Giannetti, and Banfi}}]{gandolfi2019accessing}
\bibinfo{author}{\bibfnamefont{M.}~\bibnamefont{Gandolfi}},
  \bibinfo{author}{\bibfnamefont{G.}~\bibnamefont{Benetti}},
  \bibinfo{author}{\bibfnamefont{C.}~\bibnamefont{Glorieux}},
  \bibinfo{author}{\bibfnamefont{C.}~\bibnamefont{Giannetti}},
  \bibnamefont{and} \bibinfo{author}{\bibfnamefont{F.}~\bibnamefont{Banfi}},
  \bibinfo{journal}{International Journal of Heat and Mass Transfer}
  \textbf{\bibinfo{volume}{143}}, \bibinfo{pages}{118553}
  (\bibinfo{year}{2019}).

\bibitem[{\citenamefont{Fivez et~al.}(2011)\citenamefont{Fivez, Salenbien,
  Malayil, Schols, and Glorieux}}]{fivez2011dynamics}
\bibinfo{author}{\bibfnamefont{J.}~\bibnamefont{Fivez}},
  \bibinfo{author}{\bibfnamefont{R.}~\bibnamefont{Salenbien}},
  \bibinfo{author}{\bibfnamefont{M.~K.} \bibnamefont{Malayil}},
  \bibinfo{author}{\bibfnamefont{W.}~\bibnamefont{Schols}}, \bibnamefont{and}
  \bibinfo{author}{\bibfnamefont{C.}~\bibnamefont{Glorieux}}, in
  \emph{\bibinfo{booktitle}{Journal of Physics: Conference Series}}
  (\bibinfo{organization}{IOP Publishing}, \bibinfo{year}{2011}), vol.
  \bibinfo{volume}{278}, p. \bibinfo{pages}{012021}.

\bibitem[{\citenamefont{Havriliak and Negami}(1966)}]{havriliak1966complex}
\bibinfo{author}{\bibfnamefont{S.}~\bibnamefont{Havriliak}} \bibnamefont{and}
  \bibinfo{author}{\bibfnamefont{S.}~\bibnamefont{Negami}}, in
  \emph{\bibinfo{booktitle}{Journal of Polymer Science Part C: Polymer
  Symposia}} (\bibinfo{organization}{Wiley Online Library},
  \bibinfo{year}{1966}), vol.~\bibinfo{volume}{14}, pp.
  \bibinfo{pages}{99--117}.

\bibitem[{\citenamefont{Auld}(1973)}]{auld1973acoustic}
\bibinfo{author}{\bibfnamefont{B.~A.} \bibnamefont{Auld}},
  \emph{\bibinfo{title}{Acoustic fields and waves in solids}}
  (\bibinfo{publisher}{John Wiley \& Sons}, \bibinfo{year}{1973}).

\bibitem[{\citenamefont{Gandolfi et~al.}(2020)\citenamefont{Gandolfi, Banfi,
  and Glorieux}}]{gandolfi2020optical}
\bibinfo{author}{\bibfnamefont{M.}~\bibnamefont{Gandolfi}},
  \bibinfo{author}{\bibfnamefont{F.}~\bibnamefont{Banfi}}, \bibnamefont{and}
  \bibinfo{author}{\bibfnamefont{C.}~\bibnamefont{Glorieux}},
  \bibinfo{journal}{Photoacoustics} \textbf{\bibinfo{volume}{20}},
  \bibinfo{pages}{100199} (\bibinfo{year}{2020}).

\bibitem[{\citenamefont{Mukhopadhyay}(1999)}]{mukhopadhyay1999relaxation}
\bibinfo{author}{\bibfnamefont{S.}~\bibnamefont{Mukhopadhyay}},
  \bibinfo{journal}{Journal of thermal stresses} \textbf{\bibinfo{volume}{22}},
  \bibinfo{pages}{829} (\bibinfo{year}{1999}).

\bibitem[{\citenamefont{Othman and Abbas}(2012)}]{othman2012fundamental}
\bibinfo{author}{\bibfnamefont{M.}~\bibnamefont{Othman}} \bibnamefont{and}
  \bibinfo{author}{\bibfnamefont{I.}~\bibnamefont{Abbas}},
  \bibinfo{journal}{Computational Mathematics and Modeling}
  \textbf{\bibinfo{volume}{23}}, \bibinfo{pages}{158} (\bibinfo{year}{2012}).

\bibitem[{\citenamefont{Yan and Nelson}(1987)}]{yan1987impulsive}
\bibinfo{author}{\bibfnamefont{Y.-X.} \bibnamefont{Yan}} \bibnamefont{and}
  \bibinfo{author}{\bibfnamefont{K.~A.} \bibnamefont{Nelson}},
  \bibinfo{journal}{The Journal of chemical physics}
  \textbf{\bibinfo{volume}{87}}, \bibinfo{pages}{6240} (\bibinfo{year}{1987}).

\bibitem[{\citenamefont{Gupta and Kumar}(2012)}]{gupta2012scope}
\bibinfo{author}{\bibfnamefont{M.}~\bibnamefont{Gupta}} \bibnamefont{and}
  \bibinfo{author}{\bibfnamefont{N.}~\bibnamefont{Kumar}},
  \bibinfo{journal}{Renewable and Sustainable Energy Reviews}
  \textbf{\bibinfo{volume}{16}}, \bibinfo{pages}{4551} (\bibinfo{year}{2012}).

\bibitem[{\citenamefont{Jackson}(1976)}]{jackson1976most}
\bibinfo{author}{\bibfnamefont{D.~D.} \bibnamefont{Jackson}},
  \bibinfo{journal}{Journal of Geophysical Research}
  \textbf{\bibinfo{volume}{81}}, \bibinfo{pages}{1027} (\bibinfo{year}{1976}).

\bibitem[{\citenamefont{Salenbien et~al.}(2011)\citenamefont{Salenbien, Cote,
  Goossens, Limaye, Labie, and Glorieux}}]{salenbien2011laser}
\bibinfo{author}{\bibfnamefont{R.}~\bibnamefont{Salenbien}},
  \bibinfo{author}{\bibfnamefont{R.}~\bibnamefont{Cote}},
  \bibinfo{author}{\bibfnamefont{J.}~\bibnamefont{Goossens}},
  \bibinfo{author}{\bibfnamefont{P.}~\bibnamefont{Limaye}},
  \bibinfo{author}{\bibfnamefont{R.}~\bibnamefont{Labie}}, \bibnamefont{and}
  \bibinfo{author}{\bibfnamefont{C.}~\bibnamefont{Glorieux}},
  \bibinfo{journal}{Journal of applied physics} \textbf{\bibinfo{volume}{109}},
  \bibinfo{pages}{093104} (\bibinfo{year}{2011}).

\bibitem[{\citenamefont{Caddeo et~al.}(2017)\citenamefont{Caddeo, Melis,
  Ronchi, Giannetti, Ferrini, Rurali, Colombo, and Banfi}}]{caddeo2017thermal}
\bibinfo{author}{\bibfnamefont{C.}~\bibnamefont{Caddeo}},
  \bibinfo{author}{\bibfnamefont{C.}~\bibnamefont{Melis}},
  \bibinfo{author}{\bibfnamefont{A.}~\bibnamefont{Ronchi}},
  \bibinfo{author}{\bibfnamefont{C.}~\bibnamefont{Giannetti}},
  \bibinfo{author}{\bibfnamefont{G.}~\bibnamefont{Ferrini}},
  \bibinfo{author}{\bibfnamefont{R.}~\bibnamefont{Rurali}},
  \bibinfo{author}{\bibfnamefont{L.}~\bibnamefont{Colombo}}, \bibnamefont{and}
  \bibinfo{author}{\bibfnamefont{F.}~\bibnamefont{Banfi}},
  \bibinfo{journal}{Physical Review B} \textbf{\bibinfo{volume}{95}},
  \bibinfo{pages}{085306} (\bibinfo{year}{2017}).

\bibitem[{\citenamefont{Verstraeten et~al.}(2015)\citenamefont{Verstraeten,
  Sermeus, Salenbien, Fivez, Shkerdin, and
  Glorieux}}]{verstraeten2015determination}
\bibinfo{author}{\bibfnamefont{B.}~\bibnamefont{Verstraeten}},
  \bibinfo{author}{\bibfnamefont{J.}~\bibnamefont{Sermeus}},
  \bibinfo{author}{\bibfnamefont{R.}~\bibnamefont{Salenbien}},
  \bibinfo{author}{\bibfnamefont{J.}~\bibnamefont{Fivez}},
  \bibinfo{author}{\bibfnamefont{G.}~\bibnamefont{Shkerdin}}, \bibnamefont{and}
  \bibinfo{author}{\bibfnamefont{C.}~\bibnamefont{Glorieux}},
  \bibinfo{journal}{Photoacoustics} \textbf{\bibinfo{volume}{3}},
  \bibinfo{pages}{64} (\bibinfo{year}{2015}).

\bibitem[{\citenamefont{Vega-Flick et~al.}(2015)\citenamefont{Vega-Flick,
  Eliason, Maznev, Khanolkar, Abi~Ghanem, Boechler, Alvarado-Gil, and
  Nelson}}]{vega2015laser}
\bibinfo{author}{\bibfnamefont{A.}~\bibnamefont{Vega-Flick}},
  \bibinfo{author}{\bibfnamefont{J.}~\bibnamefont{Eliason}},
  \bibinfo{author}{\bibfnamefont{A.}~\bibnamefont{Maznev}},
  \bibinfo{author}{\bibfnamefont{A.}~\bibnamefont{Khanolkar}},
  \bibinfo{author}{\bibfnamefont{M.}~\bibnamefont{Abi~Ghanem}},
  \bibinfo{author}{\bibfnamefont{N.}~\bibnamefont{Boechler}},
  \bibinfo{author}{\bibfnamefont{J.}~\bibnamefont{Alvarado-Gil}},
  \bibnamefont{and} \bibinfo{author}{\bibfnamefont{K.}~\bibnamefont{Nelson}},
  \bibinfo{journal}{Review of Scientific Instruments}
  \textbf{\bibinfo{volume}{86}}, \bibinfo{pages}{123101}
  (\bibinfo{year}{2015}).

\bibitem[{\citenamefont{Maznev et~al.}(1998)\citenamefont{Maznev, Nelson, and
  Rogers}}]{maznev1998optical}
\bibinfo{author}{\bibfnamefont{A.}~\bibnamefont{Maznev}},
  \bibinfo{author}{\bibfnamefont{K.}~\bibnamefont{Nelson}}, \bibnamefont{and}
  \bibinfo{author}{\bibfnamefont{J.}~\bibnamefont{Rogers}},
  \bibinfo{journal}{Optics letters} \textbf{\bibinfo{volume}{23}},
  \bibinfo{pages}{1319} (\bibinfo{year}{1998}).

\bibitem[{\citenamefont{Brodard and Vauthey}(2005)}]{brodard2005application}
\bibinfo{author}{\bibfnamefont{P.}~\bibnamefont{Brodard}} \bibnamefont{and}
  \bibinfo{author}{\bibfnamefont{E.}~\bibnamefont{Vauthey}},
  \bibinfo{journal}{The Journal of Physical Chemistry B}
  \textbf{\bibinfo{volume}{109}}, \bibinfo{pages}{4668} (\bibinfo{year}{2005}).

\bibitem[{\citenamefont{Taschin et~al.}(2008)\citenamefont{Taschin, Eramo,
  Bartolini, and Torre}}]{taschin2008time}
\bibinfo{author}{\bibfnamefont{A.}~\bibnamefont{Taschin}},
  \bibinfo{author}{\bibfnamefont{R.}~\bibnamefont{Eramo}},
  \bibinfo{author}{\bibfnamefont{P.}~\bibnamefont{Bartolini}},
  \bibnamefont{and} \bibinfo{author}{\bibfnamefont{R.}~\bibnamefont{Torre}},
  \emph{\bibinfo{title}{Time-resolved spectroscopy of complex liquids}}
  (\bibinfo{year}{2008}).

\bibitem[{\citenamefont{Salenbien}(2012)}]{salenbien2012photoacoustic}
\bibinfo{author}{\bibfnamefont{R.}~\bibnamefont{Salenbien}}
  (\bibinfo{year}{2012}).

\bibitem[{\citenamefont{Liu and Nagel}(1998)}]{liu1998jamming}
\bibinfo{author}{\bibfnamefont{A.~J.} \bibnamefont{Liu}} \bibnamefont{and}
  \bibinfo{author}{\bibfnamefont{S.~R.} \bibnamefont{Nagel}},
  \bibinfo{journal}{Nature} \textbf{\bibinfo{volume}{396}}, \bibinfo{pages}{21}
  (\bibinfo{year}{1998}).

\bibitem[{\citenamefont{Jeong et~al.}(1986)\citenamefont{Jeong, Nagel, and
  Bhattacharya}}]{jeong1986ultrasonic}
\bibinfo{author}{\bibfnamefont{Y.~H.} \bibnamefont{Jeong}},
  \bibinfo{author}{\bibfnamefont{S.~R.} \bibnamefont{Nagel}}, \bibnamefont{and}
  \bibinfo{author}{\bibfnamefont{S.}~\bibnamefont{Bhattacharya}},
  \bibinfo{journal}{Physical Review A} \textbf{\bibinfo{volume}{34}},
  \bibinfo{pages}{602} (\bibinfo{year}{1986}).

\bibitem[{\citenamefont{Schroyen et~al.}(2020)\citenamefont{Schroyen,
  Vlassopoulos, Van~Puyvelde, and Vermant}}]{schroyen2020bulk}
\bibinfo{author}{\bibfnamefont{B.}~\bibnamefont{Schroyen}},
  \bibinfo{author}{\bibfnamefont{D.}~\bibnamefont{Vlassopoulos}},
  \bibinfo{author}{\bibfnamefont{P.}~\bibnamefont{Van~Puyvelde}},
  \bibnamefont{and} \bibinfo{author}{\bibfnamefont{J.}~\bibnamefont{Vermant}},
  \bibinfo{journal}{Rheologica Acta} \textbf{\bibinfo{volume}{59}},
  \bibinfo{pages}{1} (\bibinfo{year}{2020}).

\bibitem[{\citenamefont{Birge}(1986)}]{birge1986specific}
\bibinfo{author}{\bibfnamefont{N.~O.} \bibnamefont{Birge}},
  \bibinfo{journal}{Physical Review B} \textbf{\bibinfo{volume}{34}},
  \bibinfo{pages}{1631} (\bibinfo{year}{1986}).

\bibitem[{\citenamefont{Wang et~al.}(2002)\citenamefont{Wang, Velikov, and
  Angell}}]{wang2002direct}
\bibinfo{author}{\bibfnamefont{L.-M.} \bibnamefont{Wang}},
  \bibinfo{author}{\bibfnamefont{V.}~\bibnamefont{Velikov}}, \bibnamefont{and}
  \bibinfo{author}{\bibfnamefont{C.}~\bibnamefont{Angell}},
  \bibinfo{journal}{The Journal of chemical physics}
  \textbf{\bibinfo{volume}{117}}, \bibinfo{pages}{10184}
  (\bibinfo{year}{2002}).

\bibitem[{\citenamefont{Minakov et~al.}(2001)\citenamefont{Minakov, Adamovsky,
  and Schick}}]{minakov2001simultaneous}
\bibinfo{author}{\bibfnamefont{A.}~\bibnamefont{Minakov}},
  \bibinfo{author}{\bibfnamefont{S.}~\bibnamefont{Adamovsky}},
  \bibnamefont{and} \bibinfo{author}{\bibfnamefont{C.}~\bibnamefont{Schick}},
  \bibinfo{journal}{Thermochimica acta} \textbf{\bibinfo{volume}{377}},
  \bibinfo{pages}{173} (\bibinfo{year}{2001}).

\bibitem[{\citenamefont{B{\"o}hmer et~al.}(1993)\citenamefont{B{\"o}hmer, Ngai,
  Angell, and Plazek}}]{bohmer1993nonexponential}
\bibinfo{author}{\bibfnamefont{R.}~\bibnamefont{B{\"o}hmer}},
  \bibinfo{author}{\bibfnamefont{K.}~\bibnamefont{Ngai}},
  \bibinfo{author}{\bibfnamefont{C.~A.} \bibnamefont{Angell}},
  \bibnamefont{and} \bibinfo{author}{\bibfnamefont{D.}~\bibnamefont{Plazek}},
  \bibinfo{journal}{The Journal of chemical physics}
  \textbf{\bibinfo{volume}{99}}, \bibinfo{pages}{4201} (\bibinfo{year}{1993}).

\end{thebibliography}

\end{document}